\documentclass[10pt]{article}
\usepackage{graphicx}
\usepackage{amsmath}
\usepackage{amssymb}
\usepackage{caption2}
\begin{document}
\bibliographystyle{prsty}
\begin{center}
{\large {\bf \sc{ Analysis of the  $T_{c\bar{s}}(2900)$ and related tetraquark states with the QCD sum rules
  }}} \\[2mm]
Xiao-Song Yang$^{*\dag}$, Qi Xin$^{*\dag}$, Zhi-Gang Wang$^*$ \footnote{E-mail: zgwang@aliyun.com.  }  \\
 Department of Physics, North China Electric Power University, Baoding 071003, P. R. China$^*$
 School of Nuclear Science and Engineering, North China Electric Power University, Beijing 102206, P. R. China$^\dag$
\end{center}

\begin{abstract}
In this research, we tentatively assign the $T_{c\bar{s}}(2900)$ as the $A\bar{A}$-type  tetraquark state,   and study the mass spectrum  of the tetraquark states with strange and doubly  strange, which have the spin-parity $J^P = 0^+$, $1^+$ and $2^+$, in the framework of the QCD sum rules in details, where the $A$ denotes the axialvector diquark state. The predicted mass $M=2.92\pm0.12\,\rm{GeV}$ is in consistent with the experimental values $M=2.892\pm0.014\pm0.015\,\rm{GeV}$ and $2.921\pm0.017\pm0.020\,\rm{GeV}$ from the LHCb collaboration and supports assigning the $T_{c\bar{s}}(2900)$ to be the $A\bar{A}$-type scalar $c\bar{s}q\bar{q}$ tetraquark state. The predictions for other tetraquark states can be confronted to the experimental data in the future to diagnose the nature of the fully open flavor exotic states.
\end{abstract}



\section{Introduction}

In 2020, the LHCb collaboration carried out the amplitude analysis of the process $B^+\to D^+ D^- K^+$ using the LHCb $pp$ collision data taken at $\sqrt{s}=7,8,$ and $13\,\rm{TeV}$ corresponding to a total integrated luminosity of 9 fb${}^{-1}$ and observed a narrow peak in the $D^- K^+$ invariant mass spectrum \cite{LHCbX2900,LHCbX2900-2}. The peak can be   parameterized in terms of two Breit-Wigner resonances with the spin-0 and spin-1, the masses and widths  are determined to be,
\begin{eqnarray}
X_0\left(2900\right)&:&M=2.866 \pm0.007 \pm0.002\, \rm{GeV}\, , \,\Gamma=57 \pm12 \pm4 \, \rm{MeV} \, , \nonumber\\
X_1\left(2900\right)&:&M=2.904 \pm0.005 \pm0.001 \, \rm{GeV}\, , \, \Gamma=110 \pm11 \pm4 \, \rm{MeV} \, ,
\end{eqnarray}
respectively. It is the first exotic structure  with fully open flavor and leads to several research works to explore its inner structures, such as the tetraquark state \cite{X2900-Lu-Tetra,WangX2900,X2900-He-Tetra,X2900-Karliner-Tetra,X2900-Zhang-Tetra,X2900-Agaev-Tetra}, molecular state \cite{X2900-Hu-Mole,X2900-Liu-Mole,X2900-Chen-Mole,X2900-Huang-Mole}, triangle singularity and cusp effect  \cite{X2900-Liu-Triangle,X2900-Burns-Triangle}. Among those interpretations, we adopted  the tetraquark state scenario,   and studied the axialvector-diquark-axialvector-antidiquark $\left(A\bar{A}\right)$ type and scalar-diquark-scalar-antidiquark $\left(S\bar{S}\right)$ type fully open flavor $cs\bar{u}\bar{d}$ tetraquark states with the spin-parity $J^P=0^+$ via the QCD sum rules, and got the conclusion  that it is reasonable to assign the $X_0\left(2900\right)$ to be the $A\bar{A}$-type $cs\bar{u}\bar{d}$ tetraquark state.

Very recently, the LHCb collaboration observed the tetraquark candidates $T^a_{c\bar{s}0}(2900)^{0/++}$ with the spin-parity  $J^P=0^+$ in the processes $B^+ \to D^- D_s^+ \pi^+$ and $B^0 \to \bar{D}^0 D_s^+ \pi^-$ with the  significance larger than $9\sigma$ \cite{LHCbTcs2900-1,LHCbTcs2900-2}. The measured Breit-Wigner masses and widths  are
\begin{eqnarray}
T^a_{c\bar{s}0}(2900)^{0}&:&M=2.892\pm0.014\pm0.015\,\rm{GeV}\, , \,\Gamma=0.119\pm0.026\pm0.013\,\rm{GeV}\, , \nonumber\\
T^a_{c\bar{s}0}(2900)^{++}&:&M=2.921\pm0.017\pm0.020\,\rm{GeV}\, , \, \Gamma=0.137\pm0.032\pm0.017\,\rm{GeV}\, ,
\end{eqnarray}
respectively, such tetraquark candidates maybe exist  in the diquark-antidiquark picture \cite{X2900-Lu-Tetra,X2900-He-Tetra,X2900-Agaev-Tetra}, due to the light-flavor $SU(3)$ symmetry and the flavor-blinded strong interaction.

They may be the isospin partners of the tetraquark candidates composed of the $c\bar{s}u\bar{d}$ and $c\bar{s}\bar{u}d$, and hence have attracted a lot of research works to comprehend their nature.
In the scenario of tetraquark states, the masses of the $T^a_{c\bar{s}0}(2900)^{0/++}$ can be reproduced in the
framework of the nonrealistic potential quark model \cite{Tcs2900-Liu-Tetra}, and the color flux-tube model \cite{Tcs2900-Wei-Tetra}. The $T_{c\bar{s}0}(2900)^{0}$, $T_{c\bar{s}0}(2900)^{++}$ and $X_0(2900)$ can be
accommodated in the light flavor $SU(3)_F$ symmetry sextet \cite{V.D2900}.
In the scenario of molecular states, the $T^a_{c\bar{s}0}(2900)^{0/++}$ can (cannot) be assigned as the $D^*K^*$ molecular states in the
framework of the QCD sum rules \cite{Tcs2900-Agaev-mole,Tcs2900-Agaev-mole-2}, the one-boson-exchange model \cite{Tcs2900-Chen-mole}, and
the effective Lagrangian approach \cite{Tcs2900-Yue-mole} (Bethe-Salpeter equation \cite{Tcs2900-Ke-Non-mole}).
On the other hand, we cannot exclude that they are not real resonances and just threshold effects \cite{threshold effects,threshold effects-Oset}.  More works on analogous structures are still needed to diagnose their nature.
(After the present work is finished, Ref.\cite{Decay-Chen} appears, where the strong decays of the $T^a_{c\bar{s}0}(2900)^{0/++}$ in the scenario of tetraquark states are studied with the QCD sum rules.)

In our previous works, we have investigated the hidden-charm, doubly-charm, hidden-bottom tetraquark (molecular) states in our special scheme  via the QCD sum rules in a comprehensive way, and reach satisfactory assignments of the exotic $X$, $Y$, $Z$ and $T$ states and made many predictions \cite{WZG3985-CPC,WZG-EPJC-cc,WZG-XQ-EPJA,WZG-HC-PRD,WZG-HB-EPJC,WZG-cc-IJMPA}. In the present work, we extend our previous works to make possible assignments  of the $T^a_{c\bar{s}0}(2900)^{0/++}$, just like what we have done  in Ref.\cite{WangX2900}. Due to the similar masses, spin-parity and quark constituents of the $T^a_{c\bar{s}0}(2900)^{0/++}$ and $X_0\left(2900\right)$, we consider to assign the $T^a_{c\bar{s}0}(2900)^{0/++}$ and $X_0\left(2900\right)$ in the same picture  as the compact $A\bar{A}$-type tetraquark states tentatively, and construct the $A\bar{A}$-type  currents to study the ground state  mass spectrum of the tetraquark states with strange and doubly  strange (and have the spin-parity $J^P = 0^+$, $1^+$ and $2^+$) via the QCD sum rules to verify the inner structures of the $T^a_{c\bar{s}0}(2900)^{0/++}$.

The article is organized as follows: we get the QCD sum rules for the tetraquark states with strange and doubly strange in Section 2; in Section 3, we present the numerical results and discussions;  finally, Section 4 is reserved for our conclusion.

\section{ QCD sum rules for the tetraquark states }

Let us write down the two-point correlation functions at first,
\begin{eqnarray}
\Pi(p)&=&i\int d^4x\, e^{ipx}\, \langle 0|T\left\{J(x)J^{\dagger}(0)\right\}|0\rangle\, , \nonumber  \\
\Pi_{\mu\nu\alpha\beta}(p)&=&i\int d^4x\, e^{ipx}\, \langle 0|T\left\{J_{\mu\nu}(x)J_{\alpha\beta}^{\dagger}(0)\right\}|0\rangle \, ,
\end{eqnarray}
where the currents $J(x)=J^0_s(x)$, $J^0_{ss}(x)$ and $J_{\mu\nu}(x)=J_{s,\mu\nu}^{1}(x)$, $J_{ss,\mu\nu}^{1}(x)$, $J_{s,\mu\nu}^{2}(x)$, $J_{ss,\mu\nu}^{2}(x)$,
\begin{eqnarray}
J^0_s(x)&=&\varepsilon^{ijk}\varepsilon^{imn}u^T_j (x)C\gamma_\mu c_k(x) \bar{d}_m(x)\gamma^\mu C \bar{s}^T_n(x) \, ,\nonumber \\
J^0_{ss}(x)&=&\varepsilon^{ijk}\varepsilon^{imn}u^T_j (x)C\gamma_\mu c_k(x) \bar{s}_m(x)\gamma^\mu C \bar{s}^T_n(x) \, ,\nonumber \\
J^1_{s,\mu\nu}(x)&=&\varepsilon^{ijk}\varepsilon^{imn}\left[u^T_j (x)C\gamma_\mu c_k(x) \bar{d}_m(x)\gamma_\nu C \bar{s}^T_n(x)-u^T_j (x)C\gamma_\nu c_k(x) \bar{d}_m(x)\gamma_\mu C \bar{s}^T_n(x)\right] \, , \nonumber \\
J^1_{ss,\mu\nu}(x)&=&\varepsilon^{ijk}\varepsilon^{imn}\left[u^T_j (x)C\gamma_\mu c_k(x) \bar{s}_m(x)\gamma_\nu C \bar{s}^T_n(x)-u^T_j (x)C\gamma_\nu c_k(x) \bar{s}_m(x)\gamma_\mu C \bar{s}^T_n(x)\right] \, , \nonumber \\
J^2_{s,\mu\nu}(x)&=&\varepsilon^{ijk}\varepsilon^{imn}\left[u^T_j (x)C\gamma_\mu c_k(x) \bar{d}_m(x)\gamma_\nu C \bar{s}^T_n(x)+u^T_j (x)C\gamma_\nu c_k(x) \bar{d}_m(x)\gamma_\mu C \bar{s}^T_n(x)\right] \, , \nonumber \\
J^2_{ss,\mu\nu}(x)&=&\varepsilon^{ijk}\varepsilon^{imn}\left[u^T_j (x)C\gamma_\mu c_k(x) \bar{s}_m(x)\gamma_\nu C \bar{s}^T_n(x)+u^T_j (x)C\gamma_\nu c_k(x) \bar{s}_m(x)\gamma_\mu C \bar{s}^T_n(x)\right] \, ,
\end{eqnarray}
the $i$, $j$, $k$, $m$ and $n$ are color indexes, (thereafter) the superscripts $0$, $1$ and $2$ denote the spins of the currents, the $\mu$ and $\nu$ are the Lorentz indices,  and the subscripts $s$ and $ss$ denote the currents with strange and doubly strange, respectively. With a simple replacement $u\leftrightarrow d$, we obtain the corresponding currents in the same isospin multiplets. In the isospin limit, the tetraquark states in the same multiplets have the same masses.

On the hadron side, we insert a complete set of intermediate hadronic states with the same quantum numbers as the currents into the correlation functions \cite{SVZ79,Reinders85}, and isolate the ground state contributions,
\begin{eqnarray}
\Pi^0(p)&=& \frac{\lambda^2_{T}}{M^2_{T}-p^2}+\cdots \, ,\nonumber\\
&=&\Pi_{T}(p^2) \,\, , \nonumber \\
\Pi^1_{\mu\nu\alpha\beta}(p)&=&\frac{\lambda^2_{T}}{M^2_{T} \left(M^2_{T}-p^2\right)}\left(p^2g_{\mu\alpha}g_{\nu\beta}- p^2g_{\mu\beta}g_{\nu\alpha}-g_{\mu\alpha}p_{\nu} p_{\beta}-g_{\nu\beta}p_{\mu} p_{\alpha}+g_{\mu\beta}p_{\nu} p_{\alpha}+g_{\nu\alpha}p_{\mu} p_{\beta}\right)  \nonumber\\
&&+\frac{\lambda^2_{Y}}{M^2_{Y} \left(M^2_{Y}-p^2\right)}\left(-g_{\mu\alpha}p_{\nu} p_{\beta}-g_{\nu\beta}p_{\mu} p_{\alpha}+g_{\mu\beta}p_{\nu} p_{\alpha}+g_{\nu\alpha}p_{\mu} p_{\beta}\right)+\cdots \, , \nonumber\\
&=&\widetilde{\Pi}_{T}(p^2)\left(p^2g_{\mu\alpha}g_{\nu\beta}- p^2g_{\mu\beta}g_{\nu\alpha}-g_{\mu\alpha}p_{\nu} p_{\beta}-g_{\nu\beta}p_{\mu} p_{\alpha}+g_{\mu\beta}p_{\nu} p_{\alpha}+g_{\nu\alpha}p_{\mu} p_{\beta}\right)  \nonumber\\
&&+\widetilde{\Pi}_{Y}(p^2)\left(-g_{\mu\alpha}p_{\nu} p_{\beta}-g_{\nu\beta}p_{\mu} p_{\alpha}+g_{\mu\beta}p_{\nu} p_{\alpha}+g_{\nu\alpha}p_{\mu} p_{\beta}\right) \, , \nonumber\\
\Pi^2_{\mu\nu\alpha\beta}(p)&=&\frac{\lambda^2_{T}}{M^2_{T}-p^2}\left(\frac
{\widetilde{g}_{\mu\alpha}\widetilde{g}_{\nu\beta}+\widetilde{g}_{\mu\beta}\widetilde{g}_{\nu\alpha}}{2}- \frac{\widetilde{g}_{\mu\nu}\widetilde{g}_{\alpha\beta}}{3}\right) +\cdots \, , \nonumber\\
&=&\Pi_{T}(p^2)\left(\frac
{\widetilde{g}_{\mu\alpha}\widetilde{g}_{\nu\beta}+\widetilde{g}_{\mu\beta}\widetilde{g}_{\nu\alpha}}{2}- \frac{\widetilde{g}_{\mu\nu}\widetilde{g}_{\alpha\beta}}{3}\right) +\cdots \, ,
\end{eqnarray}
where $\widetilde{g}_{\mu\nu}=g_{\mu\nu}-\frac{p_{\mu}p_{\nu}}{p^2}$, the pole residues are defined by
\begin{eqnarray}
\langle0|J^0(0)|T(p)\rangle&=&\lambda_{T} \,\, , \nonumber \\
\langle0|J_{\mu\nu}^{1}(0)|T(p)\rangle&=&\frac{\lambda_{T}}{M_{T}} \varepsilon_{\mu\nu\alpha\beta} \varepsilon^{\alpha} p^{\beta} \,\, , \nonumber \\
\langle0|J_{\mu\nu}^{1}(0)|Y(p)\rangle&=&\frac{\lambda_{Y}}{M_{Y}} \left(\varepsilon_{\mu}p_{\nu}-\varepsilon_{\nu}p_{\mu}\right) \,\, , \nonumber \\
\langle0|J_{\mu\nu}^{2}(0)|T(p)\rangle&=&\lambda_{T}\varepsilon_{\mu\nu} \,\, ,
\end{eqnarray}
the current $J^1_{\mu\nu}(x)$ couples potentially to both  the $J^P=1^+$ and $1^{-}$ tetraquark states $T$ and $Y$, respectively, and the $\varepsilon_{\mu}$ and $\varepsilon_{\mu\nu}$ are polarization vectors of the tetraquark states. We choose the components $\Pi_T(p^2)$ to explore the properties of the $T$-states.

We can project out the components $\Pi_T\left(p^2\right)$ and $\Pi_Y\left(p^2\right)$ via the projective operators $P^{\mu\nu\alpha\beta}_T$ and $P^{\mu\nu\alpha\beta}_Y$,

\begin{eqnarray}
\Pi_{T} \left(p^2\right)&=& p^2 \widetilde{\Pi}_{T}\left(p^2\right) = P^{\mu\nu\alpha\beta}_T \Pi^{T}_{\mu\nu\alpha\beta}(p) \,\, , \nonumber \\
\Pi_{Y} \left(p^2\right)&=& p^2 \widetilde{\Pi}_{Y}\left(p^2\right) = P^{\mu\nu\alpha\beta}_Y \Pi^{Y}_{\mu\nu\alpha\beta}(p) \,\, ,
\end{eqnarray}
where
\begin{eqnarray}
P^{\mu\nu\alpha\beta}_T &=& \frac{1}{6}\left(g^{\mu\alpha}-\frac{p^\mu p^\alpha}{p^2}\right) \left(g^{\nu\beta}-\frac{p^\nu p^\beta}{p^2}\right) \,\, , \nonumber \\
P^{\mu\nu\alpha\beta}_Y &=& \frac{1}{6}\left(g^{\mu\alpha}-\frac{p^\mu p^\alpha}{p^2}\right) \left(g^{\nu\beta}-\frac{p^\nu p^\beta}{p^2}\right)-\frac{1}{6}g^{\mu\alpha}g^{\nu\beta} \,\, .
\end{eqnarray}

On the QCD side, we firstly contract the $u$, $d$, $s$ and $c$ quark fields in the correlation functions  with the Wick's theorem, and obtain  one heavy full quark propagator and three light full quark propagators,
\begin{eqnarray}
U/D_{ij}\left(x\right) &=& \frac{i\delta_{ij} \!\not\!{x}}{2\pi^2 x^4}- \frac{\delta_{ij}\langle \bar{q}q \rangle}{12}- \frac{\delta_{ij}x^2\langle \bar{q}g_s \sigma Gq \rangle}{192}- \frac{i g_s G^a_{\alpha\beta} t^a_{ij} \left(\!\not\!{x}\sigma^{\alpha\beta}+\sigma^{\alpha\beta}\!\not\!{x}\right)}{32\pi^2 x^2} \nonumber \\
&&- \frac{\delta_{ij} x^4 \langle \bar{q}q \rangle \langle g^2_s GG \rangle}{27648} -\frac{1}{8} \langle \bar{q}_j \sigma^{\mu\nu} q_i \rangle \sigma_{\mu\nu} + \cdots  \,\, ,
\end{eqnarray}
\begin{eqnarray}
S_{ij}\left(x\right) &=& \frac{i\delta_{ij} \!\not\!{x}}{2\pi^2 x^4}- \frac{\delta_{ij}m_s}{4\pi^2 x^2}- \frac{\delta_{ij}\langle \bar{s}s \rangle}{12}+ \frac{i\delta_{ij} \!\not\!{x} m_s \langle \bar{s}s \rangle}{48}- \frac{\delta_{ij}x^2\langle \bar{s}g_s \sigma Gs \rangle}{192}+ \frac{i\delta_{ij} x^2 \!\not\!{x} m_s\langle \bar{s}g_s \sigma Gs \rangle}{1152} \nonumber \\
&&- \frac{i g_s G^a_{\alpha\beta} t^a_{ij} \left(\!\not\!{x}\sigma^{\alpha\beta}+\sigma^{\alpha\beta}\!\not\!{x}\right)}{32\pi^2 x^2}- \frac{\delta_{ij} x^4 \langle \bar{s}s \rangle \langle g^2_s GG \rangle}{27648} -\frac{1}{8} \langle \bar{s}_j \sigma^{\mu\nu} s_i \rangle \sigma_{\mu\nu} + \cdots  \,\, ,
\end{eqnarray}
\begin{eqnarray}
C_{ij}\left(x\right) &=& \frac{i}{\left(2\pi\right)^4} \int d^4 k e^{-i k \cdot x} \left\{ \frac{\delta_{ij}}{\!\not\!{k}-m_c}- \frac{g_s G^n_{\alpha\beta} t^n_{ij}}{4} \frac{\sigma^{\alpha\beta}\left(\!\not\!{k}+m_c\right) + \left(\!\not\!{k}+m_c\right) \sigma^{\alpha\beta}} {\left(k^2-m^2_c\right)^2} \right. \nonumber \\
&&\left. - \frac{g^2_s \left(t^a t^b\right)_{ij} G^a_{\alpha\beta} G^b_{\mu\nu} \left(f^{\alpha\beta\mu\nu}+ f^{\alpha\mu\beta\nu}+ f^{\alpha\mu\nu\beta}\right) }{4 \left(k^2-m^2_c\right)^5} + \cdots \right\} \,\, , \nonumber \\
f^{\alpha\beta\mu\nu} &=& \left(\!\not\!{k}+m_c\right) \gamma^{\alpha} \left(\!\not\!{k}+m_c\right) \gamma^{\beta} \left(\!\not\!{k}+m_c\right) \gamma^{\mu} \left(\!\not\!{k}+m_c\right) \gamma^{\nu} \left(\!\not\!{k}+m_c\right) \,\, ,
\end{eqnarray}
where $t^n=\frac{\lambda^n}{2}$, the $\lambda^n$ is the Gell-Mann matrix \cite{Reinders85,Pascual84,WangHuang-prd}, then we accomplish all the integrals to obtain the QCD representations. For technical details, the interested readers can consult Refs.\cite{WangHuang-prd,WangHuang-epjc,Wang2014,WZG-Penta-IJMPA}. If the heavy quark line emits a gluon and every light quark line contributes a quark-antiquark pair, we can get a quark-gluon operator $g_s G\bar{q}q\bar{q}q\bar{q}q$ of dimension 11, its vacuum expectation can be factorized into the form $\langle q\bar{q}\rangle\langle q\bar{q}\rangle\langle qg_s\sigma G\bar{q}\rangle$ by assuming the vacuum saturation, where $q=u,d$ or $s$. Therefore, we should calculate  the vacuum condensates up to  dimension 11.
In details, we take account of the vacuum condensates $\langle \bar{q}q \rangle$, $\langle \frac{\alpha_s GG}{\pi}\rangle$, $ \langle \bar{q}g_s \sigma Gq \rangle$, $\langle \bar{q}q \rangle^2$,
$\langle \bar{q}q \rangle \langle \frac{\alpha_s GG}{\pi}\rangle$, $\langle \bar{q}q \rangle \langle \bar{q}g_s \sigma Gq \rangle$, $\langle \bar{q}q \rangle^3$, $\langle \bar{q}g_s \sigma Gq \rangle^2 $, $\langle \bar{q}q \rangle^2 \langle \frac{\alpha_s GG}{\pi}\rangle$, $\langle \bar{q}q \rangle^2 \langle \bar{q}g_s \sigma Gq \rangle$,  with the truncations of the operators of the orders $\mathcal{O}\left(\alpha_s^k\right)$  with $k\leq 1$  consistently, just like  in our previous studies \cite{WangHuang-prd,WangHuang-epjc,Wang2014,WZG-Penta-IJMPA}.

We obtain the QCD spectral densities $\rho_{QCD}\left(s\right)$ through dispersion relation, and the explicit spectral densities $\rho_{QCD}\left(s\right)$ are given in the Appendix. Then, we match the hadron spectral densities  with QCD spectral densities   bellow the continuum thresholds with the assumption of  quark-hadron duality, and perform Borel transform in regard to the variable $P^2=-p^2$ to get the QCD sum rules,
\begin{eqnarray}\label{QCDsumrule}
  \lambda_{T}^2 \exp\left(-\frac{M^2_T}{T^2}\right)&=& \int_{m_c^2}^{s_0}ds \rho_{QCD}\left(s\right) \exp\left(-\frac{s}{T^2}\right) \, ,
\end{eqnarray}
where the $T^2$ is the Borel parameter.

Finally, we differentiate Eq.\eqref{QCDsumrule} with respect to $\tau=\frac{1}{T^2}$ and obtain the masses of the tetraquark states,
\begin{eqnarray}
M_T^2 &=& -\frac{\frac{d}{d\tau}\int_{m_c^2}^{s_0}ds\, \rho_{QCD}\left(s\right) \exp\left(-\tau s\right)}{\int_{m_c^2}^{s_0}ds \,\rho_{QCD}\left(s\right) \exp\left(-\tau s\right)} \, .
\end{eqnarray}

\section{Numerical results and discussions}	
We adopt the standard values of the input parameters on the QCD side
$\langle\bar{q}q \rangle=-(0.24\pm 0.01\, \rm{GeV})^3$,
$\langle\bar{s}s \rangle=(0.8 \pm 0.1) \langle\bar{q}q \rangle$,
 $\langle\bar{q}g_s\sigma G q \rangle=m_0^2\langle \bar{q}q \rangle$,
 $\langle\bar{s}g_s\sigma G s \rangle=m_0^2\langle \bar{s}s \rangle$,
$m_0^2=(0.8 \pm 0.1)\,\rm{GeV}^2$,  $\langle \frac{\alpha_{s}GG}{\pi} \rangle=(0.33\, \rm{GeV})^4$  at the energy scale  $\mu=1\, \rm{GeV}$
\cite{SVZ79,Reinders85,Colangelo-Review}, and take the modified minimal subtraction masses of the charm
quark and strange quark $m_{c}(m_c)=(1.275\pm0.025)\,\rm{GeV}$ and $m_{s}(\mu=2\,\rm{GeV})=(0.095\pm0.005)\,\rm{GeV}$
 from the Particle Data Group \cite{PDG}. Furthermore, we set $m_u=m_d=0$ because of  the small current quark masses and take account of the energy-scale dependence of the input parameters from the renormalization group equation \cite{Narison-mix},
\begin{eqnarray}
\langle\bar{q}q \rangle(\mu)&=&\langle\bar{q}q \rangle(1\, \rm{GeV}) \left[\frac{\alpha_{s}(1\, \rm{GeV})}{\alpha_{s}(\mu)}\right]^{\frac{12}{33-2n_f}} \, , \nonumber\\
\langle\bar{s}s \rangle(\mu)&=&\langle\bar{s}s \rangle(1\, \rm{GeV}) \left[\frac{\alpha_{s}(1\, \rm{GeV})}{\alpha_{s}(\mu)}\right]^{\frac{12}{33-2n_f}} \, , \nonumber\\
\langle\bar{q}g_s \sigma Gq \rangle(\mu)&=&\langle\bar{q}g_s \sigma Gq \rangle(1\, \rm{GeV})\left[\frac{\alpha_{s}(1\, \rm{GeV})}{\alpha_{s}(\mu)}\right]^{\frac{2}{33-2n_f}}\, , \nonumber\\
\langle\bar{s}g_s \sigma Gs \rangle(\mu)&=&\langle\bar{s}g_s \sigma Gs \rangle(1\, \rm{GeV})\left[\frac{\alpha_{s}(1\, \rm{GeV})}{\alpha_{s}(\mu)}\right]^{\frac{2}{33-2n_f}}\, , \nonumber\\
m_c(\mu)&=&m_c(m_c)\left[\frac{\alpha_{s}(\mu)}{\alpha_{s}(m_c)}\right]^{\frac{12}{33-2n_f}} \, ,\nonumber\\
m_s(\mu)&=&m_s(2\, \rm{GeV})\left[\frac{\alpha_{s}(\mu)}{\alpha_{s}(2\, \rm{GeV})}\right]^{\frac{12}{33-2n_f}} \, ,\nonumber\\
\alpha_s(\mu)&=&\frac{1}{b_0t}\left[1-\frac{b_1}{b_0^2}\frac{\log t}{t} +\frac{b_1^2(\log^2{t}-\log{t}-1)+b_0b_2}{b_0^4t^2}\right]\, ,
\end{eqnarray}
where $t=\log \frac{\mu^2}{\Lambda^2}$, $b_0=\frac{33-2n_f}{12\pi}$, $b_1=\frac{153-19n_f}{24\pi^2}$, $b_2=\frac{2857-\frac{5033}{9}n_f+\frac{325}{27}n_f^2}{128\pi^3}$,  $\Lambda=213\,\rm{MeV}$, $296\,\rm{MeV}$  and  $339\,\rm{MeV}$ for the flavors  $n_f=5$, $4$ and $3$, respectively  \cite{PDG}. We take the flavor numbers $n_f=4$ (i.e. $u$, $d$, $s$ and $c$) and the typical energy scale $\mu=1.0\,\rm{GeV}$ in the present study as in our previous works \cite{WangX2900}.

The energy gaps from the ground states to the first radial excited states are $0.5\sim0.7\,\rm{GeV}$ approximately for the charmed mesons. We tentatively adjust the continuum threshold parameters  $\sqrt{s_0}=M_T+0.5\sim0.7\,\rm{GeV}$ and Borel parameters $T^2$, and  search for the best values via trial and error to avoid the contaminations from the higher resonances and continuum states, and to satisfy the two fundamental criteria of the QCD sum rules, which are pole dominance at the hadron side and convergence of the operator product expansion at the QCD side. The pole contributions are defined by
\begin{eqnarray}
\rm{pole} &=& \frac{\int^{s_0}_{m^2_c} ds \rho_{QCD}\left(s\right)\exp\left(-\frac{s}{T^2}\right)}{\int^{\infty}_{m^2_c} ds \rho_{QCD}\left(s\right)\exp\left(-\frac{s}{T^2}\right)} \, ,
\end{eqnarray}
and are required to reach the values about $40\% - 60\%$ consistently, just like in our previous works \cite{WangX2900,WZG3985-CPC,WZG-EPJC-cc,WZG-XQ-EPJA,WZG-HC-PRD,WZG-HB-EPJC,WZG-cc-IJMPA}.

In calculations, we determine the continuum threshold parameters and Borel parameters  by the pole contributions alone, then we check whether or not the operator product expansion is convergent automatically.
In Table \ref{mass}, we can see clearly that the maximum ranges  of the pole contributions are $38\% - 66\%$ and all the central values for  the six states are larger than $50\%$, which implies that the criterion of pole dominance is satisfied pretty well.

\begin{figure}
 \centering
  \includegraphics[totalheight=7cm,width=10cm]{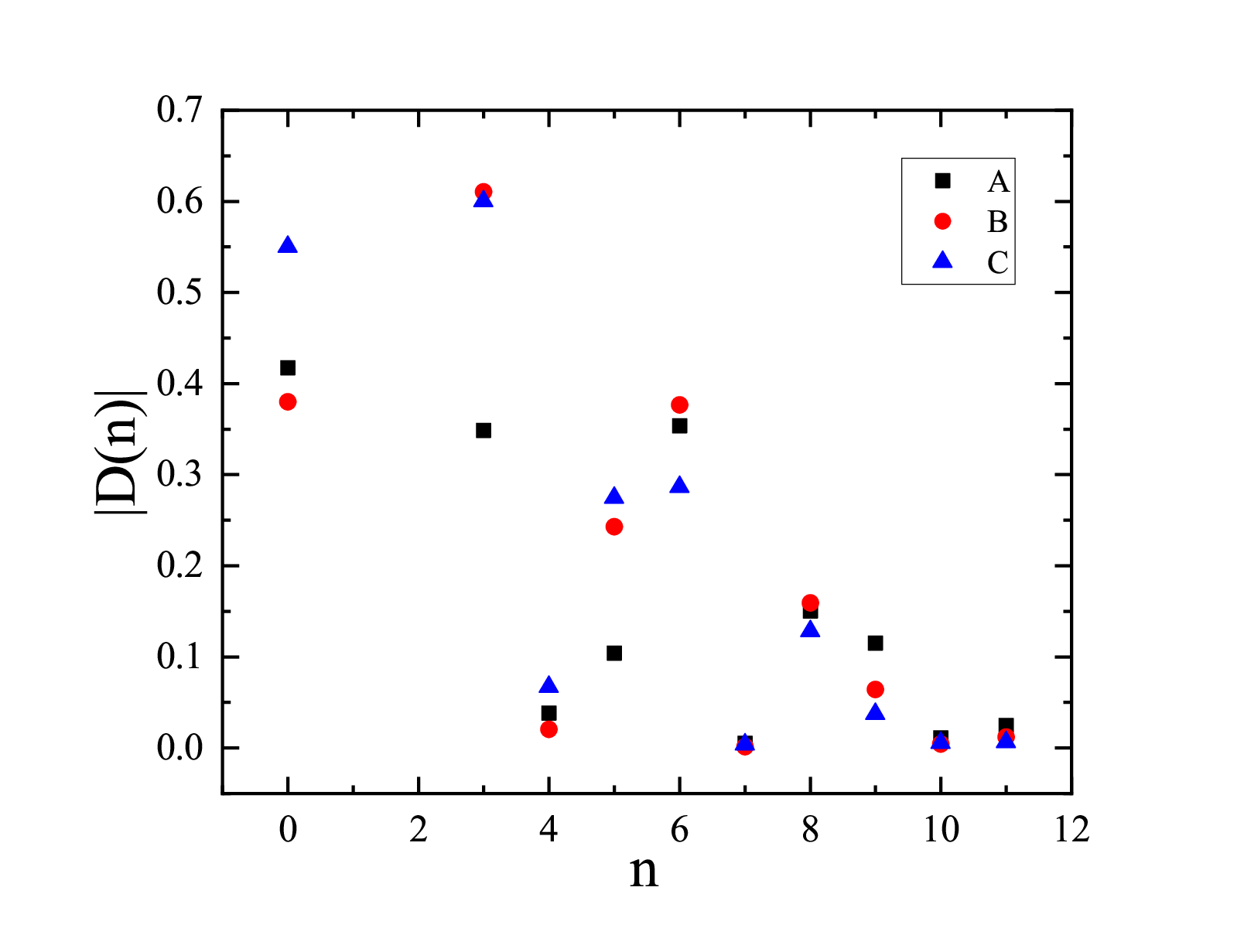}
  \caption{The contributions  $D\left(n\right)$ for the $c\bar{s}q\bar{q}$ tetraquark states, where $A$, $B$ and $C$ denote the tetraquark states with $J^P=0^+$, $1^+$ and $2^+$, respectively.}\label{dn}
\end{figure}

In addition, we calculate the contributions of the vacuum condensates in regard to  dimension, which is defined by
\begin{eqnarray}
D\left(n\right)&=&\frac{\int_{m_c^2}^{s_0}ds \,\rho_{n,QCD}\left(s\right) \exp\left(-\frac{s}{T^2}\right)}{\int_{m_c^2}^{s_0}ds \,\rho_{QCD}\left(s\right) \exp\left(-\frac{s}{T^2}\right)} \, ,
\end{eqnarray}
to judge the convergence of the operator product expansion, where the $n$ presents the spectral densities concerning the condensates of dimension $n$. We plot the results  of the $c\bar{s}q\bar{q}$ type tetraquark states as an example in Fig.\ref{dn}, and find that the contributions $D(0)$, $D(3)$, $D(5)$, $D(6)$ and $D(8)$ play  a major role, while the contributions  $D(4)$ and $D(7)$ are rather small due to the tiny contributions of the gluon condensate $\langle \frac{\alpha_s GG}{\pi}\rangle$. The absolute values  show a descending trend  $|D(6)|>|D(8)|>|D(9)|>|D(10)|\sim |D(11)|\sim 0$. The criterion of convergence of the operator product expansion is satisfied very well.

\begin{figure}
 \centering
 \includegraphics[totalheight=5cm,width=7cm]{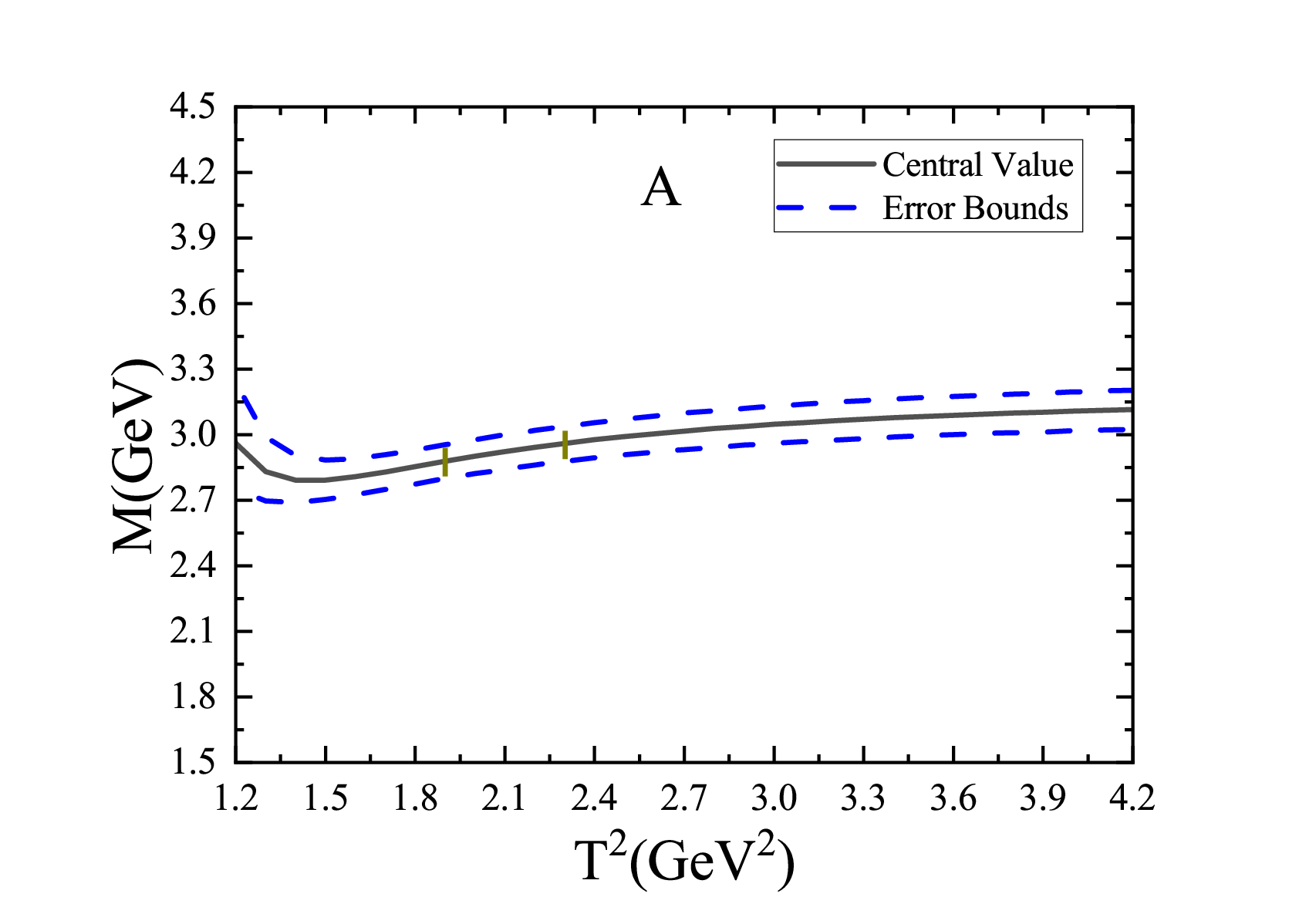}
 \includegraphics[totalheight=5cm,width=7cm]{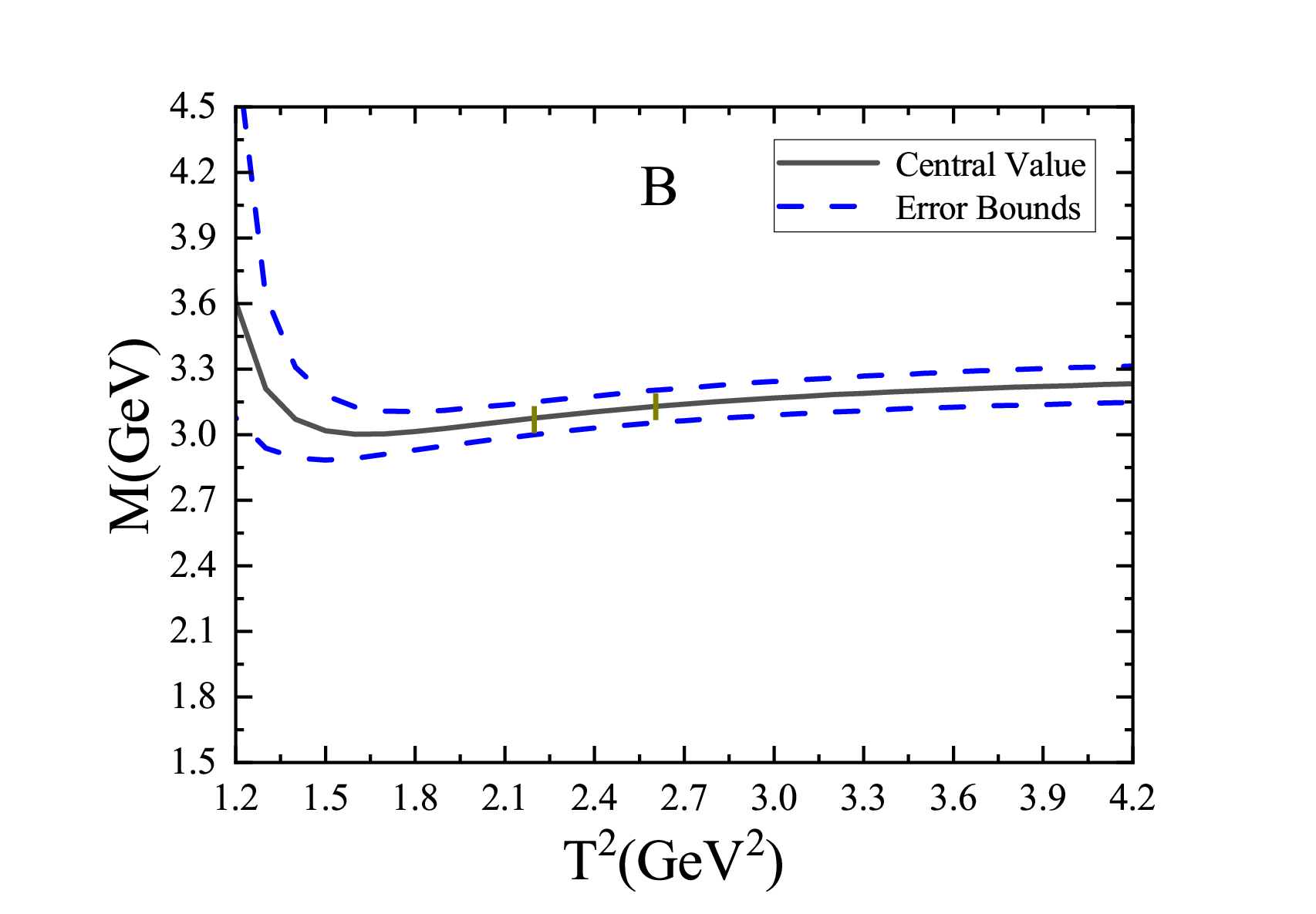}
 \includegraphics[totalheight=5cm,width=7cm]{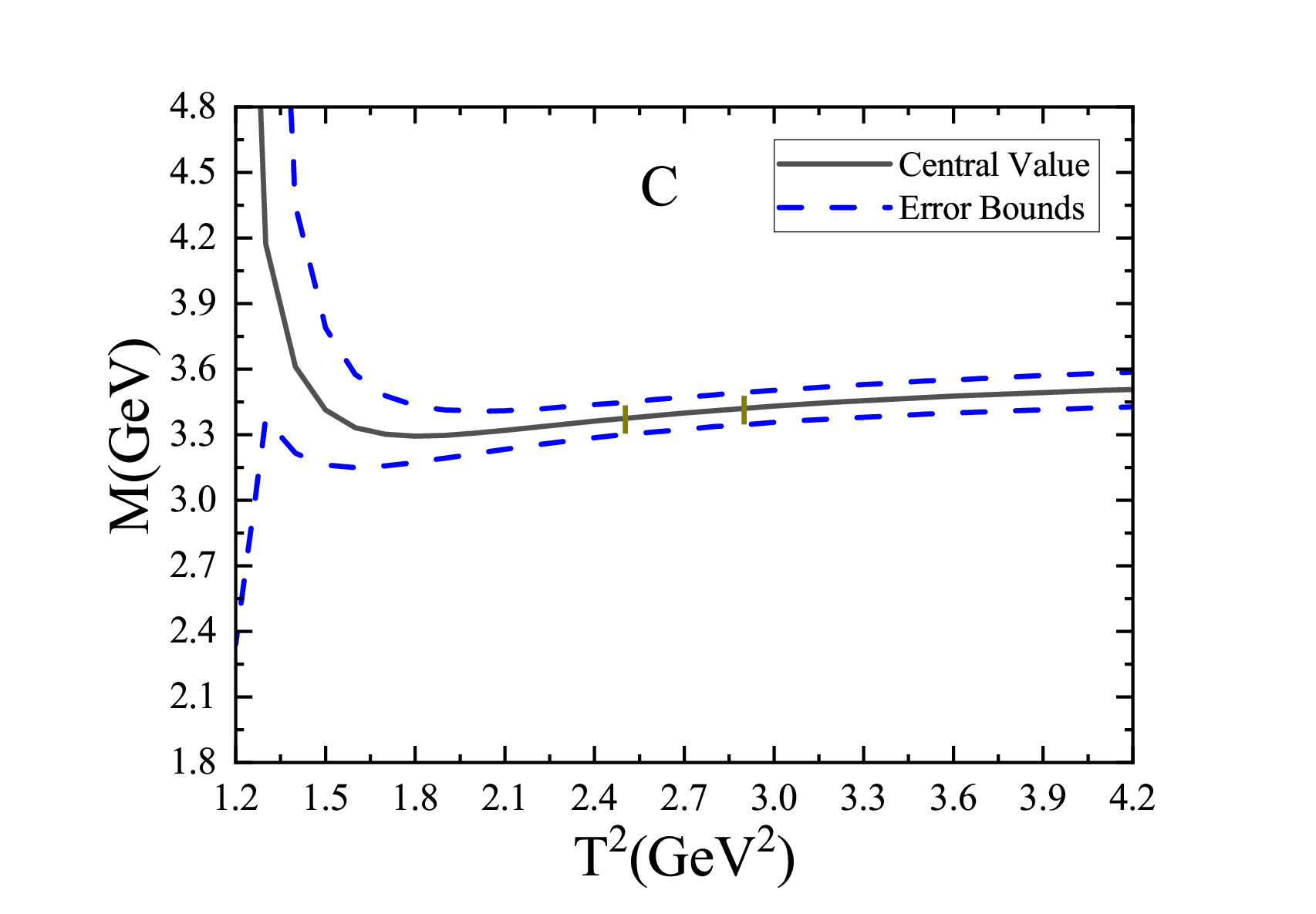}
 \includegraphics[totalheight=5cm,width=7cm]{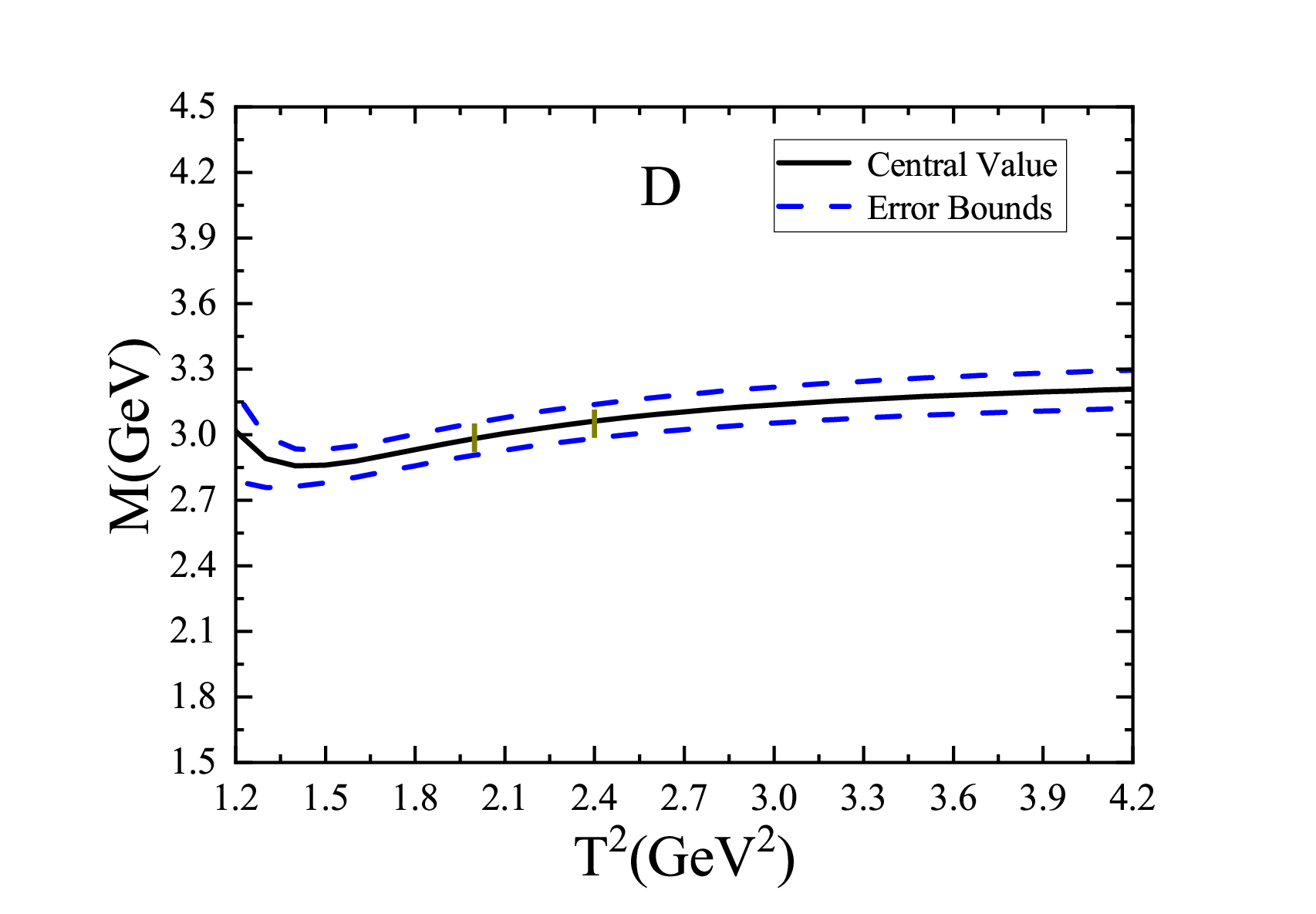}
 \includegraphics[totalheight=5cm,width=7cm]{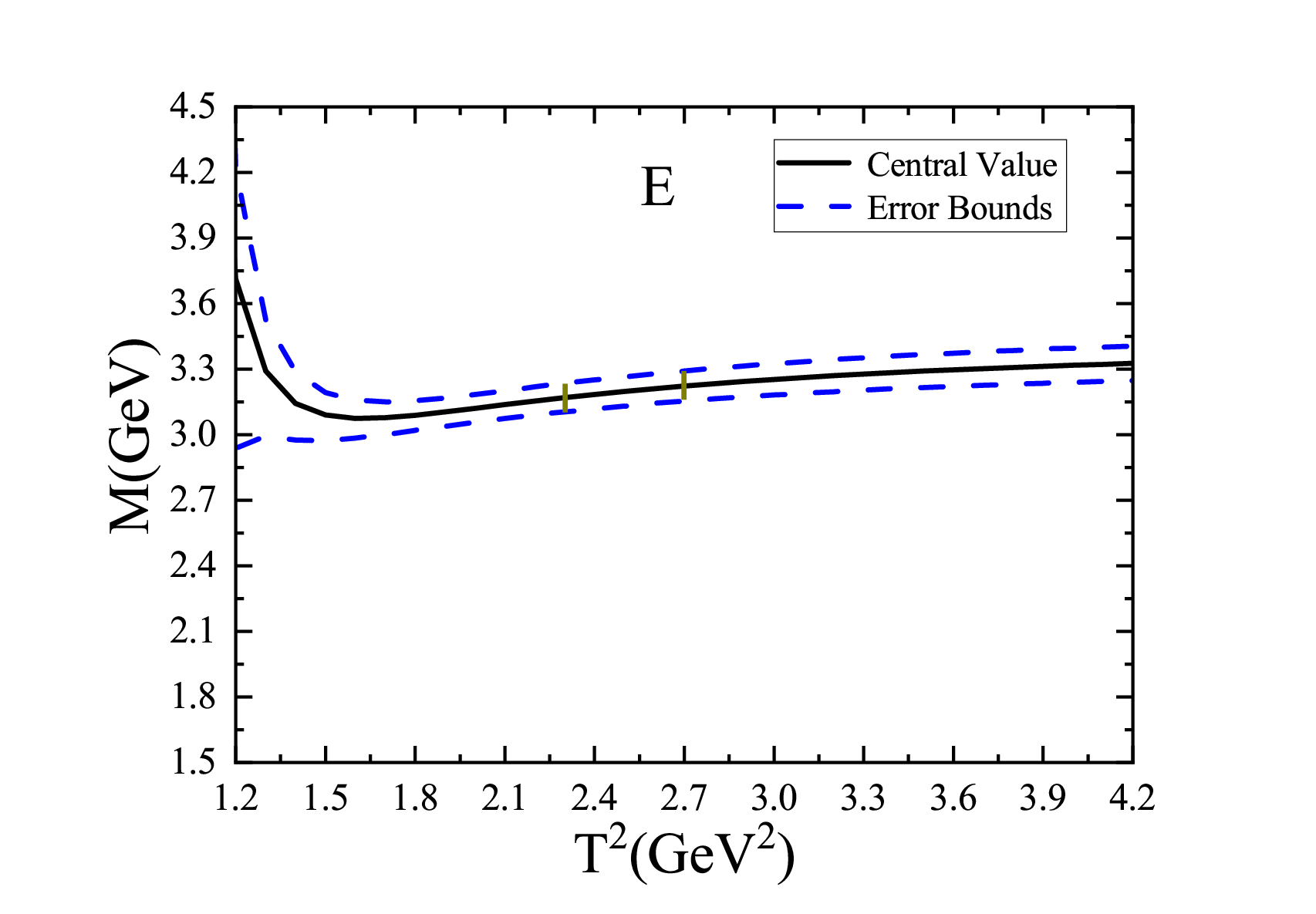}
 \includegraphics[totalheight=5cm,width=7cm]{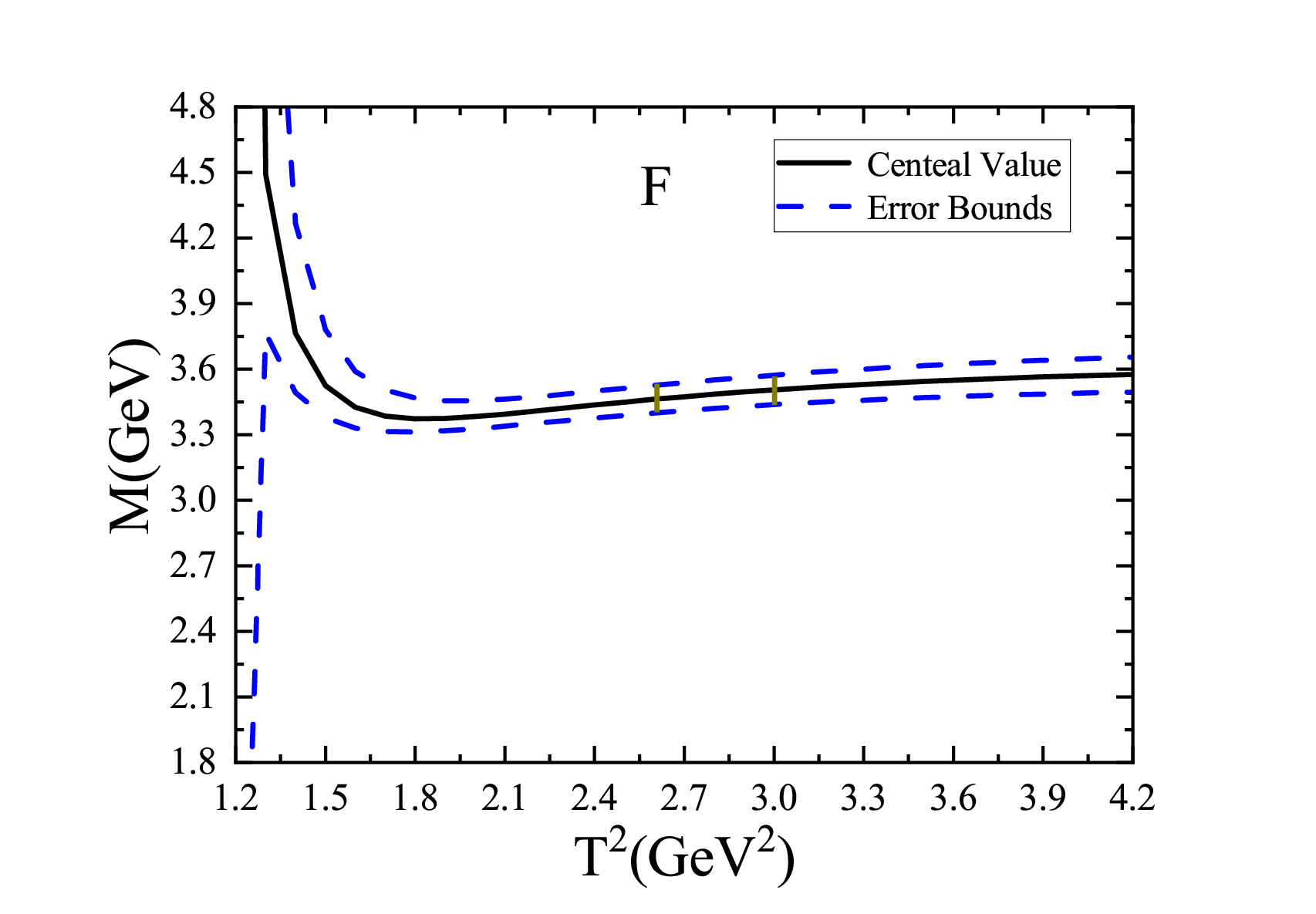}
  \caption{The masses with variations of the Borel parameters $T^2$, where the $A$, $B$ and $C$ ($D$, $E$ and $F$) denote the $J^P=0^+$, $1^+$ and $2^+$ tetraquark states with strange (doubly strange), respectively. }\label{Borel-mass}
\end{figure}

Then we consider all the uncertainties of the input parameters to get the values of the masses and pole residues,  and plot the masses of the six tetraquark states with respect to the Borel parameters $T^2$ in
Fig.\ref{Borel-mass}. We can see clearly that there appear flat platforms in the Borel windows, which are the regions between the two short perpendicular  lines, for all the six states, and thus it declares the reliability of our predictions.

In Table \ref{mass}, we display the spin-parity, Borel parameters, continuum threshold parameters, pole contributions, masses and pole residues of the tetraquark states with strange and doubly strange. The predicted mass of the $J^P=0^+$ state $cu\bar{d}\bar{s}$, $M = 2.92\pm0.12\,\rm{GeV}$, is in very good agreement with the experimental values  $M=2.892\pm0.014\pm0.015\,\rm{GeV}$ and $2.921\pm0.017\pm0.020\,\rm{GeV}$ from the LHCb collaboration \cite{LHCbTcs2900-1,LHCbTcs2900-2}, and supports assigning the $T_{c\bar{s}}(2900)$ to be the $A\bar{A}$-type $c\bar{s}q\bar{q}$ tetraquark states with the spin-parity  $J^P=0^+$. The pole residues can be used to calculate the decay widths of the corresponding tetraquark states in the following works. As the experimental data for the fully open flavor exotic states are still far from complete at the present time, we expect for further experimental data to confront with our predictions for the other related tetraquark states to illuminate the nature of the fully open flavor exotic hadrons.

\begin{table}
\begin{center}
\begin{tabular}{|c|c|c|c|c|c|c|c|}\hline\hline
                   &$J^P$    &$T^2 (\rm{GeV}^2)$  &$\sqrt{s_0}(\rm GeV) $  &pole        &$M_T(\rm{GeV})$  &$\lambda_T(10^{-2}\rm{GeV}^5)$ \\ \hline

$cu\bar{d}\bar{s}$ &$0^+$    &$1.9-2.3$           &$3.50\pm0.10$           &$(40-65)\%$  &$2.92\pm0.12$   &$1.62\pm0.33 $  \\ \hline

$cu\bar{d}\bar{s}$ &$1^+$    &$2.2-2.6$           &$3.65\pm0.10$           &$(40-62)\%$  &$3.10\pm0.10$   &$1.59\pm0.27$  \\ \hline

$cu\bar{d}\bar{s}$ &$2^+$    &$2.5-2.9$           &$3.95\pm0.10$           &$(42-61)\%$  &$3.40\pm0.10$   &$3.51\pm0.55$  \\ \hline

$cu\bar{s}\bar{s}$ &$0^+$    &$2.0-2.4$           &$3.60\pm0.10$           &$(38-66)\%$  &$3.02\pm0.12$   &$2.72\pm0.43$  \\ \hline

$cu\bar{s}\bar{s}$ &$1^+$    &$2.3-2.7$           &$3.75\pm0.10$           &$(40-65)\%$  &$3.20\pm0.10$   &$2.64\pm0.33$  \\ \hline

$cu\bar{s}\bar{s}$ &$2^+$    &$2.6-3.0$           &$4.05\pm0.10$           &$(41-64)\%$  &$3.49\pm0.10$   &$5.70\pm0.65$  \\ \hline
\end{tabular}
\end{center}
\caption{ The spin-parity, Borel parameters, continuum threshold parameters, pole contributions, masses and pole residues of the tetraquark states. }\label{mass}
\end{table}

\section{Conclusion}

In this article, we construct the $A\bar{A}$-type  currents to study the fully open heavy flavor tetraquark states with strange and doubly strange which have the spin-parity $J^P = 0^+$, $1^+$ and $2^+$ via the QCD sum rules approach in details. We perform the operator product expansion up to the vacuum condensates of dimension 11 in a consistent way, and obtain the masses and pole residues of those tetraquark states. The predicted mass of the $cu\bar{d}\bar{s}$ state, $M=2.92\pm0.12\,\rm{GeV}$, is in very good agreement with the experimental data $M=2.892\pm0.014\pm0.015\,\rm{GeV}$ and $2.921\pm0.017\pm0.020\,\rm{GeV}$ from the LHCb collaboration,
  as a consequence, it is reasonable to interpret the $T_{c\bar{s}}(2900)$ as the $A\bar{A}$-type scalar $cu\bar{d}\bar{s}$ and $cd\bar{u}\bar{s}$ tetraquark states. Besides, we  expect more experimental data to examine our predictions for the other tetraquark states in the future to illustrate the nature of the $T$-states.

\section*{Appendix}
The detailed QCD spectral densities,
 \begin{eqnarray}
 \rho_{QCD}(s)&=&\rho_{s/ss}^{0/1/2}\left(n\right)\, ,
 \end{eqnarray}
 where the superscripts  $0$, $1$ and $2$ stand for the spins of the tetraquark states, the subscripts $s$ and $ss$ stand for the valence $s$-quarks, and the $n$ stands for the dimensions of the condensates,

\begin{eqnarray}
\rho^0_s\left(0\right)&=& \frac{1}{1536\pi^6}\int_{x_i}^1 dx x\left(1-x\right)^4 \left(3s-\tilde{m}_c^2\right) \left(s-\tilde{m}_c^2\right)^3\nonumber \, ,
\end{eqnarray}

\begin{eqnarray}
\rho^0_s\left(3\right) &=&-\frac{m_s\left(\langle\bar{q}q\rangle-\langle\bar{s}s\rangle\right)}{16\pi^4} \int_{x_i}^1 dx x\left(1-x\right)^2 \left(2s-\tilde{m}_c^2\right)\left(s-\tilde{m}_c^2\right)\nonumber\\
&&-\frac{m_c\langle\bar{q}q\rangle}{32\pi^4} \int_{x_i}^1 dx \left(1-x\right)^2 \left(s-\tilde{m}_c^2\right)^2\nonumber\, ,
\end{eqnarray}

\begin{eqnarray}
\rho^0_s\left(4\right)&=&\langle\frac{\alpha_{s}GG}{\pi}\rangle \Big\{ \frac{1}{768\pi^4}\int_{x_i}^1 dx \left(1+2x\right)\left(1-x\right)^2 \left(2s-\tilde{m}_c^2\right)\left(s-\tilde{m}_c^2\right)\nonumber \\
&&-\frac{m_c^2}{2304\pi^4}\int_{x_i}^1 dx \frac{\left(1-x\right)^4}{x^2}\left(3s-2\tilde{m}_c^2\right) \Big\}\nonumber\, ,
\end{eqnarray}

\begin{eqnarray}
\rho^0_s\left(5\right) &=&\frac{m_s\left(9\langle\bar{q}g_s\sigma Gq\rangle-8\langle\bar{s}g_s\sigma Gs\rangle\right)}{384\pi^4} \int_{x_i}^1 dx x\left(1-x\right) \left(3s-2\tilde{m}_c^2\right)\nonumber\\
&&-\frac{m_s\langle\bar{q}g_s\sigma Gq\rangle}{256\pi^4} \int_{x_i}^1 dx \left(1-x\right)^2 \left(3s-2\tilde{m}_c^2\right) +\frac{m_c\langle\bar{q}g_s\sigma Gq\rangle}{64\pi^4} \int_{x_i}^1 dx \left(1-x\right) \left(s-\tilde{m}_c^2\right)\nonumber\, ,
\end{eqnarray}

\begin{eqnarray}
\rho^0_s\left(6\right) &=&\frac{\langle\bar{q}q\rangle \langle\bar{s}s\rangle}{6\pi^2} \int_{x_i}^1 dx x\left(1-x\right) \left(3s-2\tilde{m}_c^2\right) +\frac{m_s m_c\langle\bar{q}q\rangle\left(4\langle\bar{q}q\rangle-\langle\bar{s}s\rangle\right)}{12\pi^2} \int_{x_i}^1 dx \nonumber\, ,
\end{eqnarray}

\begin{eqnarray}
\rho^0_s\left(7\right) &=&\langle\frac{\alpha_{s}GG}{\pi}\rangle \Big\{\frac{m_s m_c^2\left(\langle\bar{q}q\rangle-\langle\bar{s}s\rangle\right)}{288\pi^2}\int_{x_i}^1 dx \frac{\left(1-x\right)^2}{x^2} \left(1+\frac{s}{T^2}\right)\delta\left(s-\tilde{m}_c^2\right) \nonumber\\
&&+\frac{m_c^3\langle\bar{q}q\rangle}{288\pi^2}\int_{x_i}^1 dx \frac{\left(1-x\right)^2}{x^3}\delta\left(s-\tilde{m}_c^2\right) +\frac{m_c\langle\bar{q}q\rangle} {288\pi^2}\int_{x_i}^1 dx \left(2-\frac{3\left(1-x\right)}{x^2}\right)\nonumber\\
&&+\frac{m_s\left(4\langle\bar{q}q\rangle+\langle\bar{s}s\rangle\right)}{384\pi^2}\int_{x_i}^1 dx \left(1-x\right) \left[2+s\delta\left(s-\tilde{m}_c^2\right)\right]  \nonumber\\
&& -\frac{m_s\left(4\langle\bar{q}q\rangle-3\langle\bar{s}s\rangle\right)}{1152\pi^2}\int_{x_i}^1 dx x \left[2+s\delta\left(s-\tilde{m}_c^2\right)\right]\Big\}\nonumber\, ,
\end{eqnarray}

\begin{eqnarray}
\rho^0_s\left(8\right) &=&\frac{\langle\bar{q}q\rangle \langle\bar{s}g_s\sigma Gs\rangle+\langle\bar{s}s\rangle \langle\bar{q}g_s\sigma Gq\rangle}{96\pi^2} \int_{x_i}^1 dx \left(1-4x\right) \left[2+s\delta\left(s-\tilde{m}_c^2\right)\right] \nonumber\\
&&+\frac{m_s m_c \left(8\langle\bar{q}q\rangle\langle\bar{s}g_s\sigma Gs\rangle+9\langle\bar{s}s\rangle \langle\bar{q}g_s\sigma Gq\rangle -96\langle\bar{q}q\rangle\langle\bar{q}g_s\sigma Gq\rangle\right)}{576\pi^2} \delta\left(s-m_c^2\right)\nonumber\\
&&+\frac{m_s m_c \langle\bar{q}q\rangle\langle\bar{q}g_s\sigma Gq\rangle}{48\pi^2} \int_{x_i}^1 dx \frac{1}{x} \delta\left(s-\tilde{m}_c^2\right)\nonumber\, ,
\end{eqnarray}

\begin{eqnarray}
\rho^0_s\left(9\right) &=&-\frac{4m_c\langle\bar{q}q\rangle^2 \langle\bar{s}s\rangle}{9} \delta\left(s-m_c^2\right)\nonumber\, ,
\end{eqnarray}

\begin{eqnarray}
\rho^0_s\left(10\right) &=&\frac{\langle\bar{q}g_s\sigma Gq\rangle\langle\bar{s}g_s\sigma Gs\rangle}{1536\pi^2} \left\{8\left(1+\frac{s}{T^2}\right)\delta\left(s-m_c^2\right)+3\int_{x_i}^1 dx \left(1+\frac{s}{T^2}\right) \delta\left(s-\tilde{m}_c^2\right)\right\}\nonumber\\
&&+\frac{m_s m_c^3 \langle\bar{q}g_s\sigma Gq\rangle \left(8\langle\bar{q}g_s\sigma Gq\rangle- \langle\bar{s}g_s\sigma Gs\rangle\right)}{384\pi^2 T^4}\delta\left(s-m_c^2\right) -\frac{m_s m_c \langle\bar{q}g_s\sigma Gq\rangle^2}{192\pi^2 T^2} \delta\left(s-m_c^2\right)\nonumber\\
&&+\langle\frac{\alpha_{s}GG}{\pi}\rangle \left\{ \frac{\langle\bar{q}q\rangle\langle\bar{s}s\rangle}{216} \left[-3\int_{x_i}^1 dx\left(1+\frac{s}{T^2}\right) \delta\left(s-\tilde{m}_c^2\right)+2\left(1+\frac{s}{T^2}\right) \delta\left(s-m_c^2\right)\right] \right.\nonumber\\
&&-\frac{m_s m_c^3 \langle\bar{q}q\rangle \left(4\langle\bar{q}q\rangle-\langle\bar{s}s\rangle\right)}{216T^4} \int_{x_i}^1 dx \frac{1}{x^3} \delta\left(s-\tilde{m}_c^2\right)
 +\frac{m_s m_c^3 \langle\bar{q}q\rangle \left(8\langle\bar{q}q\rangle-\langle\bar{s}s\rangle\right)}{432T^4} \delta\left(s-m_c^2\right)\nonumber\\
&&-\frac{m_s m_c \langle\bar{q}q\rangle\langle\bar{s}s\rangle}{288T^2}\delta\left(s-m_c^2\right)+\frac{m_s m_c \langle\bar{q}q\rangle \left(4\langle\bar{q}q\rangle-\langle\bar{s}s\rangle\right)}{72T^2} \int_{x_i}^1 dx \frac{1}{x^2} \delta\left(s-\tilde{m}_c^2\right)\nonumber\\
&&\left.-\frac{m_c^2\langle\bar{q}q\rangle\langle\bar{s}s\rangle}{108T^4}\int_{x_i}^1 dx \frac{s\left(1-x\right)}{x^2} \delta\left(s-\tilde{m}_c^2\right) \right\}\nonumber\, ,
\end{eqnarray}

\begin{eqnarray}
\rho^0_s\left(11\right) &=&\frac{m_c^3 \langle\bar{q}q\rangle \left(\langle\bar{q}q\rangle \langle\bar{s}g_s\sigma Gs\rangle+2\langle\bar{s}s\rangle \langle\bar{q}g_s\sigma Gq\rangle\right)}{9T^4}\delta\left(s-m_c^2\right)
\nonumber\\
&&-\frac{m_c \langle\bar{q}q\rangle \left(\langle\bar{q}q\rangle \langle\bar{s}g_s\sigma Gs\rangle+\langle\bar{s}s\rangle \langle\bar{q}g_s\sigma Gq\rangle\right)}{36T^2}\delta\left(s-m_c^2\right)\nonumber\, ,
\end{eqnarray}

\begin{eqnarray}
\rho^1_s\left(0\right) &=&\frac{1}{30720\pi^6}\int_{x_i}^1 dx x\left(1-x\right)^5 \left[\left(s-\tilde{m}_c^2\right)^4+12s\left(s-\tilde{m}_c^2\right)^3+8s^2\left(s-\tilde{m}_c^2\right)^2\right] \nonumber\\
&&+\frac{1}{18432\pi^6}\int_{x_i}^1 dx x\left(1-x\right)^4 \left(3s+\tilde{m}_c^2\right) \left(s-\tilde{m}_c^2\right)^3 \nonumber\, ,
\end{eqnarray}

\begin{eqnarray}
\rho^1_s\left(3\right) &=&-\frac{m_c\langle\bar{q}q\rangle}{288\pi^4} \int_{x_i}^1 dx \left[3\left(1-x\right)^2 \left(s-\tilde{m}_c^2\right)^2 +\left(1-x\right)^3 \left(5s-\tilde{m}_c^2\right) \left(s-\tilde{m}_c^2\right) \right]
\nonumber\\
&&-\frac{m_s \langle\bar{q}q\rangle}{32\pi^4} \int_{x_i}^1 dx x\left(1-x\right)^2 \left(3s-\tilde{m}_c^2\right)\left(s-\tilde{m}_c^2\right)\nonumber\\
&&+\frac{m_s \langle\bar{s}s\rangle}{576\pi^4} \int_{x_i}^1 dx \left\{3x\left(1-x\right)^2 \left(s^2-\tilde{m}_c^4\right) +x\left(1-x\right)^3 \left[3\left(s-\tilde{m}_c^2\right)^2+18s\left(s-\tilde{m}_c^2\right)+4s^2\right] \right\}\nonumber\, ,
\end{eqnarray}

\begin{eqnarray}
\rho^1_s\left(4\right) &=&\langle\frac{\alpha_{s}GG}{\pi}\rangle \left\{ -\frac{1}{1152\pi^4}\int_{x_i}^1 dx x \left(1-x\right)^2 \left(2s-\tilde{m}_c^2\right)\left(s-\tilde{m}_c^2\right)\nonumber \right.\\
&&+\frac{1}{13824\pi^4}\int_{x_i}^1 dx x \left(1-x\right)^3 \left[3\left(s-\tilde{m}_c^2\right)^2+18s\left(s-\tilde{m}_c^2\right)+4s^2\right]\nonumber\\
&&-\frac{1}{55296\pi^4}\int_{x_i}^1 dx \left(1-x\right)^3 \left[13\left(s-\tilde{m}_c^2\right)^2+4s\left(s-\tilde{m}_c^2\right)\right]\nonumber\\
&&-\frac{1}{221184\pi^4}\int_{x_i}^1 dx \left(1-x\right)^4 \left[9\left(s-\tilde{m}_c^2\right)^2+48s\left(s-\tilde{m}_c^2\right)+8s^2\right]\nonumber\\
&&-\frac{m_c^2}{69120\pi^4}\int_{x_i}^1 dx \frac{\left(1-x\right)^5}{x^2} \left[3\left(4s-\tilde{m}_c^2\right)+2s^2 \delta\left(s-\tilde{m}_c^2\right)\right] \nonumber\\
&&\left.-\frac{m_c^2}{13824\pi^4}\int_{x_i}^1 dx \frac{\left(1-x\right)^4}{x^2} \tilde{m}_c^2 \right\}\nonumber\, ,
\end{eqnarray}

\begin{eqnarray}
\rho^1_s\left(5\right) &=&-\frac{m_s\langle\bar{s}g_s\sigma Gs\rangle}{576\pi^4} \int_{x_i}^1 dx \left\{2x\left(1-x\right) \tilde{m}_c^2 +x\left(1-x\right)^2 \left[12s-3\tilde{m}_c^2+2s^2\delta\left(s-\tilde{m}_c^2\right)\right] \right\}\nonumber\\
&&+\frac{5m_c\langle\bar{q}g_s\sigma Gq\rangle}{1152\pi^4} \int_{x_i}^1 dx \left[2\left(1-x\right) \left(s-\tilde{m}_c^2\right) +\left(1-x\right)^2 \left(3s-\tilde{m}_c^2\right) \right]\nonumber\\
&&+\frac{m_s\langle\bar{q}g_s\sigma Gq\rangle}{768\pi^4} \int_{x_i}^1 dx \left(23x-1\right)\left(1-x\right) \left(2s-\tilde{m}_c^2\right)\nonumber\, ,
\end{eqnarray}

\begin{eqnarray}
\rho^1_s\left(6\right) &=&\frac{\langle\bar{q}q\rangle \langle\bar{s}s\rangle}{6\pi^2} \int_{x_i}^1 dx x\left(1-x\right) \left(2s-\tilde{m}_c^2\right) +\frac{m_s m_c\langle\bar{q}q\rangle\left(6\langle\bar{q}q\rangle-\langle\bar{s}s\rangle\right)}{36\pi^2} \int_{x_i}^1 dx \nonumber\\
&&-\frac{m_s m_c\langle\bar{q}q\rangle \langle\bar{s}s\rangle}{36\pi^2} \int_{x_i}^1 dx\left(1-x\right) \left[1+2s\delta\left(s-\tilde{m}_c^2\right)\right] \nonumber\, ,
\end{eqnarray}

\begin{eqnarray}
\rho^1_s\left(7\right) &=&\langle\frac{\alpha_{s}GG}{\pi}\rangle \Bigg\{-\frac{m_s \langle\bar{s}s\rangle}{13824\pi^2}\int_{x_i}^1 dx \Big\{2\left(1-x\right) \left[17+4s \delta\left(s-\tilde{m}_c^2\right)\right] \nonumber\\
&&+\left(1-x\right)^2 \left[15+2\left(13s+\frac{4s^2}{T^2}\right) \delta\left(s-\tilde{m}_c^2\right)\right] \Big\}\nonumber\\
&&+\frac{m_c\langle\bar{q}q\rangle} {6912\pi^2}\int_{x_i}^1 dx \left[32-\frac{\left(21x^2-31x+32\right)\left(1-x\right)}{x^2} -\frac{2s\left(5x^2-13x+24\right)\left(1-x\right)}{x^2}\delta\left(s-\tilde{m}_c^2\right)\right] \nonumber\\
&&+\frac{m_s m_c^2 \langle\bar{s}s\rangle}{5184\pi^2} \int_{x_i}^1 dx \left[\frac{\left(1-x\right)^2}{x^2} \left(6-\frac{3s}{T^2}\right) +\frac{\left(1-x\right)^3}{x^2} \left(2-\frac{s}{T^2}+\frac{2s^2}{T^4}\right)  \right] \delta\left(s-\tilde{m}_c^2\right) \nonumber\\
&& +\frac{m_c^3\langle\bar{q}q\rangle}{2592\pi^2}\int_{x_i}^1 dx \left[\frac{\left(2+x\right)\left(1-x\right)^2}{x^3} +\frac{2s\left(1-x\right)^3}{x^3 T^2} \right] \delta\left(s-\tilde{m}_c^2\right)\nonumber\\
&&-\frac{m_s\left(4\langle\bar{q}q\rangle-\langle\bar{s}s\rangle\right)}{1152\pi^2}\int_{x_i}^1 dx x \left[1+s \delta\left(s-\tilde{m}_c^2\right)\right] +\frac{m_s m_c^2 \langle\bar{q}q\rangle}{288\pi^2 T^2} \int_{x_i}^1 dx \frac{s\left(1-x\right)^2}{x^2}  \delta\left(s-\tilde{m}_c^2\right)\nonumber\\
&&+\frac{5m_s \langle\bar{q}q\rangle}{1152\pi^2}\int_{x_i}^1 dx \left(1-x\right) \left[3+s \delta\left(s-\tilde{m}_c^2\right)\right]  \Bigg\}\nonumber\, ,
\end{eqnarray}

\begin{eqnarray}
\rho^1_s\left(8\right) &=&\frac{\langle\bar{q}q\rangle \langle\bar{s}g_s\sigma Gs\rangle+\langle\bar{s}s\rangle \langle\bar{q}g_s\sigma Gq\rangle}{288\pi^2} \int_{x_i}^1 dx \left(1-12x\right) \left[1+s\delta\left(s-\tilde{m}_c^2\right)\right] \nonumber\\
&&+\frac{m_s m_c \left(8\langle\bar{q}q\rangle\langle\bar{s}g_s\sigma Gs\rangle+11\langle\bar{s}s\rangle \langle\bar{q}g_s\sigma Gq\rangle -144\langle\bar{q}q\rangle\langle\bar{q}g_s\sigma Gq\rangle\right)}{1728\pi^2} \delta\left(s-m_c^2\right)\nonumber\\
&&+\frac{m_s m_c \left(8\langle\bar{q}q\rangle\langle\bar{s}g_s\sigma Gs\rangle+11\langle\bar{s}s\rangle \langle\bar{q}g_s\sigma Gq\rangle \right)}{1728\pi^2} \int_{x_i}^1 dx\left(-1+\frac{2s}{T^2}\right)\delta\left(s-\tilde{m}_c^2\right) \nonumber\\
&&+\frac{m_s m_c \langle\bar{q}q\rangle\langle\bar{q}g_s\sigma Gq\rangle}{288\pi^2} \int_{x_i}^1 dx \frac{1}{x} \delta\left(s-\tilde{m}_c^2\right)\nonumber\, ,
\end{eqnarray}

\begin{eqnarray}
\rho^1_s\left(9\right) &=&-\frac{2m_c\langle\bar{q}q\rangle^2 \langle\bar{s}s\rangle}{9} \delta\left(s-m_c^2\right)\nonumber\, ,
\end{eqnarray}

\begin{eqnarray}
\rho^1_s\left(10\right) &=&\frac{\langle\bar{q}g_s\sigma Gq\rangle\langle\bar{s}g_s\sigma Gs\rangle}{4608\pi^2 T^2} \left[40s \delta\left(s-m_c^2\right)-19\int_{x_i}^1 dx s \delta\left(s-\tilde{m}_c^2\right)\right]\nonumber\\
&&+\frac{m_s m_c^3 \langle\bar{q}g_s\sigma Gq\rangle \left(36\langle\bar{q}g_s\sigma Gq\rangle- 11\langle\bar{s}g_s\sigma Gs\rangle\right)}{3456\pi^2 T^4}\delta\left(s-m_c^2\right)\nonumber\\
&&-\frac{m_s m_c \langle\bar{q}g_s\sigma Gq\rangle \left(3\langle\bar{q}g_s\sigma Gq\rangle- 11\langle\bar{s}g_s\sigma Gs\rangle\right)}{3456\pi^2 T^2} \delta\left(s-m_c^2\right)\nonumber\\
&&+\langle\frac{\alpha_{s}GG}{\pi}\rangle \left\{ \frac{\langle\bar{q}q\rangle\langle\bar{s}s\rangle}{864} \left[-5\int_{x_i}^1 dx\left(2+\frac{s}{T^2}\right) \delta\left(s-\tilde{m}_c^2\right)+\frac{8s}{T^2} \delta\left(s-m_c^2\right)\right] \right.\nonumber\\
&& +\frac{m_s m_c^3 \langle\bar{q}q\rangle \left(4\langle\bar{q}q\rangle-\langle\bar{s}s\rangle\right)}{432T^4} \delta\left(s-m_c^2\right)+\frac{m_s m_c^3 \langle\bar{q}q\rangle \langle\bar{s}s\rangle}{648} \int_{x_i}^1 dx \frac{1-x}{x^3} \left(-\frac{5}{T^4}+\frac{2s}{T^6}\right) \delta\left(s-\tilde{m}_c^2\right)\nonumber\\
&& -\frac{m_s m_c^3 \langle\bar{q}q\rangle \left(6\langle\bar{q}q\rangle-\langle\bar{s}s\rangle\right)}{648T^4} \int_{x_i}^1 dx \frac{1}{x^3} \delta\left(s-\tilde{m}_c^2\right)+\frac{m_s m_c \langle\bar{q}q\rangle \left(6\langle\bar{q}q\rangle-\langle\bar{s}s\rangle\right)}{216T^2} \int_{x_i}^1 dx \frac{1}{x^2} \delta\left(s-\tilde{m}_c^2\right)\nonumber\\
&&+\frac{11m_s m_c \langle\bar{q}q\rangle\langle\bar{s}s\rangle}{5184T^2}\delta\left(s-m_c^2\right)+\frac{m_s m_c \langle\bar{q}q\rangle \langle\bar{s}s\rangle}{5184} \int_{x_i}^1 dx \frac{24-23x}{x^2} \left(\frac{3}{T^2}-\frac{2s}{T^4}\right) \delta\left(s-\tilde{m}_c^2\right) \nonumber\\
&&\nonumber\\
&&\left.+\frac{m_c^2\langle\bar{q}q\rangle\langle\bar{s}s\rangle}{108}\int_{x_i}^1 dx \frac{1-x}{x^2} \left(\frac{1}{T^2}-\frac{s}{T^4} \right) \delta\left(s-\tilde{m}_c^2\right) \right\}\nonumber\, ,
\end{eqnarray}

\begin{eqnarray}
\rho^1_s\left(11\right) &=&\frac{m_c^3 \langle\bar{q}q\rangle \left(\langle\bar{q}q\rangle \langle\bar{s}g_s\sigma Gs\rangle+2\langle\bar{s}s\rangle \langle\bar{q}g_s\sigma Gq\rangle\right)}{18T^4}\delta\left(s-m_c^2\right)\nonumber\\
&&-\frac{m_c \langle\bar{q}q\rangle \left(\langle\bar{q}q\rangle \langle\bar{s}g_s\sigma Gs\rangle+\langle\bar{s}s\rangle \langle\bar{q}g_s\sigma Gq\rangle\right)}{216T^2}\delta\left(s-m_c^2\right)\nonumber\, ,
\end{eqnarray}

\begin{eqnarray}
\rho^2_s\left(0\right) &=& \frac{1}{46080\pi^6}\int_{x_i}^1 dx x\left(1-x\right)^4 \left(27s-7\tilde{m}_c^2\right) \left(s-\tilde{m}_c^2\right)^3\nonumber\\
&&+\frac{1}{76800\pi^6}\int_{x_i}^1 dx x\left(1-x\right)^5 \left[3\left(s-\tilde{m}_c^2\right)^4 +44s\left(s-\tilde{m}_c^2\right)^3+40s^2\left(s-\tilde{m}_c^2\right)^2\right] \nonumber\, ,
\end{eqnarray}

\begin{eqnarray}
\rho^2_s\left(3\right) &=&-\frac{m_c\langle\bar{q}q\rangle}{144\pi^4} \int_{x_i}^1 dx \left[3\left(1-x\right)^2 \left(s-\tilde{m}_c^2\right)^2 +\left(1-x\right)^3 \left(5s-\tilde{m}_c^2\right) \left(s-\tilde{m}_c^2\right) \right]\nonumber\\
&&+\frac{m_s \langle\bar{s}s\rangle}{1440\pi^4} \int_{x_i}^1 dx \left\{3x\left(1-x\right)^2 \left(17s-7\tilde{m}_c^2\right) \left(s-\tilde{m}_c^2\right)\right.\nonumber\\
&&\left.+x\left(1-x\right)^3 \left[9\left(s-\tilde{m}_c^2\right)^2 +66s\left(s-\tilde{m}_c^2\right)+20s^2\right] \right\}\nonumber\\
&&-\frac{m_s \langle\bar{q}q\rangle}{16\pi^4} \int_{x_i}^1 dx x\left(1-x\right)^2 \left(3s-\tilde{m}_c^2\right)\left(s-\tilde{m}_c^2\right)\nonumber\, ,
\end{eqnarray}

\begin{eqnarray}
\rho^2_s\left(4\right) &=&\langle\frac{\alpha_{s}GG}{\pi}\rangle \Big\{ -\frac{1}{5760\pi^4}\int_{x_i}^1 dx x \left(1-x\right)^2 \left(59s-19\tilde{m}_c^2\right)\left(s-\tilde{m}_c^2\right)\nonumber \\
&&+\frac{1}{34560\pi^4}\int_{x_i}^1 dx x \left(1-x\right)^3 \left[9\left(s-\tilde{m}_c^2\right)^2 +66s\left(s-\tilde{m}_c^2\right)+20s^2\right]\nonumber\\
&&-\frac{1}{34560\pi^4}\int_{x_i}^1 dx \left(1-x\right)^3 \left(43s-23\tilde{m}_c^2\right) \left(s-\tilde{m}_c^2\right) \nonumber\\
&&-\frac{1}{34560\pi^4}\int_{x_i}^1 dx \left(1-x\right)^4 \left[9\left(s-\tilde{m}_c^2\right)^2 +51s\left(s-\tilde{m}_c^2\right)+10s^2\right]\nonumber\\
&&-\frac{m_c^2}{172800\pi^4}\int_{x_i}^1 dx \frac{\left(1-x\right)^5}{x^2} \left[3\left(14s-3\tilde{m}_c^2\right)+10s^2 \delta\left(s-\tilde{m}_c^2\right)\right] \nonumber\\
&&-\frac{m_c^2}{34560\pi^4}\int_{x_i}^1 dx \frac{\left(1-x\right)^4}{x^2} \left(12s-7\tilde{m}_c^2\right) \Big\}\nonumber\, ,
\end{eqnarray}

\begin{eqnarray}
\rho^2_s\left(5\right) &=&\frac{7m_c\langle\bar{q}g_s\sigma Gq\rangle}{576\pi^4} \int_{x_i}^1 dx \left[2\left(1-x\right) \left(s-\tilde{m}_c^2\right) +\left(1-x\right)^2 \left(3s-\tilde{m}_c^2\right) \right]
\nonumber\\
&&-\frac{m_s\langle\bar{s}g_s\sigma Gs\rangle}{1440\pi^4} \int_{x_i}^1 dx \left\{2x\left(1-x\right) \left(12s-7\tilde{m}_c^2\right) +x\left(1-x\right)^2 \left[42s-9\tilde{m}_c^2+10s^2\delta\left(s-\tilde{m}_c^2\right)\right] \right\}\nonumber\\
&&+\frac{m_s\langle\bar{q}g_s\sigma Gq\rangle}{384\pi^4} \int_{x_i}^1 dx \left(25x+1\right)\left(1-x\right) \left(2s-\tilde{m}_c^2\right)\nonumber\, ,
\end{eqnarray}

\begin{eqnarray}
\rho^2_s\left(6\right) &=& \frac{\langle\bar{q}q\rangle \langle\bar{s}s\rangle}{3\pi^2} \int_{x_i}^1 dx x\left(1-x\right) \left(2s-\tilde{m}_c^2\right) +\frac{m_s m_c\langle\bar{q}q\rangle\left(6\langle\bar{q}q\rangle-\langle\bar{s}s\rangle\right)}{18\pi^2} \int_{x_i}^1 dx \nonumber\\
&&-\frac{m_s m_c\langle\bar{q}q\rangle \langle\bar{s}s\rangle}{18\pi^2} \int_{x_i}^1 dx\left(1-x\right) \left[1+2s\delta\left(s-\tilde{m}_c^2\right)\right] \nonumber\, ,
\end{eqnarray}

\begin{eqnarray}
\rho^2_s\left(7\right) &=&\langle\frac{\alpha_{s}GG}{\pi}\rangle \Bigg\{\frac{m_c^3\langle\bar{q}q\rangle}{1296\pi^2}\int_{x_i}^1 dx \left[\frac{\left(2+x\right)\left(1-x\right)^2}{x^3} +\frac{2s\left(1-x\right)^3}{x^3 T^2} \right] \delta\left(s-\tilde{m}_c^2\right)
\nonumber\\
&&+\frac{m_c\langle\bar{q}q\rangle} {864\pi^2}\int_{x_i}^1 dx \left[8-\frac{\left(7x^2-13x+8\right) \left(1-x\right)}{x^2} -\frac{2s\left(7x^2-5x+2\right) \left(1-x\right) }{x^2} \delta\left(s-\tilde{m}_c^2\right) \right]\nonumber\\
&&-\frac{m_s \langle\bar{s}s\rangle}{5760\pi^2}\int_{x_i}^1 dx \left\{19x^2-52x+43 + \big\{\left[10+33\left(1-x\right)^2\right]s+\frac{10s^2}{T^2} \big\}\delta\left(s-\tilde{m}_c^2\right) \right\}\nonumber\\
&&+\frac{m_s \langle\bar{q}q\rangle}{144\pi^2}\int_{x_i}^1 dx \left[3-4x+\left(1-2x\right) s \delta\left(s-\tilde{m}_c^2\right)\right]\nonumber\\
&&-\frac{m_s m_c^2 \langle\bar{s}s\rangle}{12960\pi^2} \int_{x_i}^1 dx \frac{\left(1-x\right)^2}{x^2}\left[2\left(1+2x\right) + \frac{s\left(8+7x\right)}{T^2} +\frac{10s^2\left(1-x\right)}{T^4}  \right] \delta\left(s-\tilde{m}_c^2\right)\nonumber\\
&&+\frac{m_s m_c^2 \langle\bar{q}q\rangle}{144\pi^2 T^2} \int_{x_i}^1 dx \frac{s\left(1-x\right)^2}{x^2}  \delta\left(s-\tilde{m}_c^2\right)  \Bigg\}\nonumber\, ,
\end{eqnarray}

\begin{eqnarray}
\rho^2_s\left(8\right) &=&-\frac{\langle\bar{q}q\rangle \langle\bar{s}g_s\sigma Gs\rangle+\langle\bar{s}s\rangle \langle\bar{q}g_s\sigma Gq\rangle}{144\pi^2} \int_{x_i}^1 dx \left(1+12x\right) \left[1+s\delta\left(s-\tilde{m}_c^2\right)\right] \nonumber\\
&&+\frac{m_s m_c \left(8\langle\bar{q}q\rangle\langle\bar{s}g_s\sigma Gs\rangle+13\langle\bar{s}s\rangle \langle\bar{q}g_s\sigma Gq\rangle -144\langle\bar{q}q\rangle\langle\bar{q}g_s\sigma Gq\rangle\right)}{864\pi^2} \delta\left(s-m_c^2\right)\nonumber\\
&&+\frac{m_s m_c \left(8\langle\bar{q}q\rangle\langle\bar{s}g_s\sigma Gs\rangle+13\langle\bar{s}s\rangle \langle\bar{q}g_s\sigma Gq\rangle \right)}{864\pi^2} \int_{x_i}^1 dx\left(-1+\frac{2s}{T^2}\right)\delta\left(s-\tilde{m}_c^2\right)\nonumber\\
&&-\frac{m_s m_c \langle\bar{q}q\rangle\langle\bar{q}g_s\sigma Gq\rangle}{144\pi^2} \int_{x_i}^1 dx \frac{1}{x} \delta\left(s-\tilde{m}_c^2\right)\nonumber\, ,
\end{eqnarray}

\begin{eqnarray}
\rho^2_s\left(9\right) &=&-\frac{4m_c\langle\bar{q}q\rangle^2 \langle\bar{s}s\rangle}{9} \delta\left(s-m_c^2\right)\nonumber\, ,
\end{eqnarray}

\begin{eqnarray}
\rho^2_s\left(10\right) &=&\frac{7\langle\bar{q}g_s\sigma Gq\rangle\langle\bar{s}g_s\sigma Gs\rangle}{6912\pi^2 T^2} \left[24s \delta\left(s-m_c^2\right) +5\int_{x_i}^1 dx s \delta\left(s-\tilde{m}_c^2\right)\right]\nonumber\\
&&+\frac{m_s m_c^3 \langle\bar{q}g_s\sigma Gq\rangle \left(36\langle\bar{q}g_s\sigma Gq\rangle- 13\langle\bar{s}g_s\sigma Gs\rangle\right)}{1728\pi^2 T^4}\delta\left(s-m_c^2\right)\nonumber\\
&&+\frac{m_s m_c \langle\bar{q}g_s\sigma Gq\rangle \left(3\langle\bar{q}g_s\sigma Gq\rangle+ 13\langle\bar{s}g_s\sigma Gs\rangle\right)}{1728\pi^2 T^2} \delta\left(s-m_c^2\right)\nonumber\\
&&+\langle\frac{\alpha_{s}GG}{\pi}\rangle \left\{ \frac{\langle\bar{q}q\rangle\langle\bar{s}s\rangle}{108} \left[-\int_{x_i}^1 dx\left(2+\frac{s}{T^2}\right) \delta\left(s-\tilde{m}_c^2\right)+\frac{2s}{T^2} \delta\left(s-m_c^2\right)\right] \right.\nonumber\\
&&+\frac{7m_s m_c \langle\bar{q}q\rangle\langle\bar{s}s\rangle}{1296T^2}\delta\left(s-m_c^2\right) +\frac{m_s m_c^3 \langle\bar{q}q\rangle \left(4\langle\bar{q}q\rangle-\langle\bar{s}s\rangle\right)}{216T^4} \delta\left(s-m_c^2\right)\nonumber\\
&&+\frac{m_s m_c \langle\bar{q}q\rangle \left(6\langle\bar{q}q\rangle-\langle\bar{s}s\rangle\right)}{108T^2} \int_{x_i}^1 dx \frac{1}{x^2} \delta\left(s-\tilde{m}_c^2\right)\nonumber\\
&&-\frac{m_s m_c^3 \langle\bar{q}q\rangle \left(6\langle\bar{q}q\rangle-\langle\bar{s}s\rangle\right)}{324T^4} \int_{x_i}^1 dx \frac{1}{x^3} \delta\left(s-\tilde{m}_c^2\right) \nonumber\\
&&+\frac{m_s m_c^3 \langle\bar{q}q\rangle \langle\bar{s}s\rangle}{324} \int_{x_i}^1 dx \frac{1-x}{x^3} \left(-\frac{5}{T^4}+\frac{2s}{T^6}\right) \delta\left(s-\tilde{m}_c^2\right) \nonumber\\
&&+\frac{m_s m_c \langle\bar{q}q\rangle \langle\bar{s}s\rangle}{1296} \int_{x_i}^1 dx \frac{12-13x}{x^2} \left(\frac{3}{T^2}-\frac{2s}{T^4}\right) \delta\left(s-\tilde{m}_c^2\right)\nonumber\\
&&\left.+\frac{m_c^2\langle\bar{q}q\rangle\langle\bar{s}s\rangle}{54}\int_{x_i}^1 dx \frac{1-x}{x^2} \left(\frac{1}{T^2}-\frac{s}{T^4} \right) \delta\left(s-\tilde{m}_c^2\right) \right\}\nonumber\, ,
\end{eqnarray}

\begin{eqnarray}
\rho^2_s\left(11\right) &=&\frac{m_c \langle\bar{q}q\rangle \left(\langle\bar{q}q\rangle \langle\bar{s}g_s\sigma Gs\rangle+\langle\bar{s}s\rangle \langle\bar{q}g_s\sigma Gq\rangle\right)}{108T^2}\delta\left(s-m_c^2\right)
\nonumber\\
&&+\frac{m_c^3 \langle\bar{q}q\rangle \left(\langle\bar{q}q\rangle \langle\bar{s}g_s\sigma Gs\rangle+2\langle\bar{s}s\rangle \langle\bar{q}g_s\sigma Gq\rangle\right)}{9T^4}\delta\left(s-m_c^2\right)\nonumber\, ,
\end{eqnarray}

\begin{eqnarray}
\rho^0_{ss}\left(0\right) &=& \frac{1}{1536\pi^6}\int_{x_i}^1 dx x\left(1-x\right)^4 \left(3s-\tilde{m}_c^2\right) \left(s-\tilde{m}_c^2\right)^3\nonumber\, ,
\end{eqnarray}

\begin{eqnarray}
\rho^0_{ss}\left(3\right) &=&-\frac{m_c\langle\bar{q}q\rangle}{32\pi^4} \int_{x_i}^1 dx \left(1-x\right)^2 \left(s-\tilde{m}_c^2\right)^2\nonumber\, ,
\end{eqnarray}

\begin{eqnarray}
\rho^0_{ss}\left(4\right) &=&\langle\frac{\alpha_{s}GG}{\pi}\rangle \Big\{ \frac{1}{768\pi^4}\int_{x_i}^1 dx \left(1+2x\right)\left(1-x\right)^2 \left(2s-\tilde{m}_c^2\right)\left(s-\tilde{m}_c^2\right)\nonumber \\
&&-\frac{m_c^2}{2304\pi^4}\int_{x_i}^1 dx \frac{\left(1-x\right)^4}{x^2}\left(3s-2\tilde{m}_c^2\right) \Big\}\nonumber\, ,
\end{eqnarray}

\begin{eqnarray}
\rho^0_{ss}\left(5\right) &=&\frac{m_s\langle\bar{s}g_s\sigma Gs\rangle}{192\pi^4} \int_{x_i}^1 dx x\left(1-x\right) \left(3s-2\tilde{m}_c^2\right) -\frac{m_s\langle\bar{s}g_s\sigma Gs\rangle}{128\pi^4} \int_{x_i}^1 dx \left(1-x\right)^2 \left(3s-2\tilde{m}_c^2\right)\nonumber\\
&&+\frac{m_c\langle\bar{q}g_s\sigma Gq\rangle}{64\pi^4} \int_{x_i}^1 dx \left(1-x\right) \left(s-\tilde{m}_c^2\right)\nonumber\, ,
\end{eqnarray}

\begin{eqnarray}
\rho^0_{ss}\left(6\right) &=&\frac{\langle\bar{s}s\rangle^2}{6\pi^2} \int_{x_i}^1 dx x\left(1-x\right) \left(3s-2\tilde{m}_c^2\right)+\frac{m_s m_c\langle\bar{q}q\rangle\langle\bar{s}s\rangle}{2\pi^2} \int_{x_i}^1 dx \nonumber\, ,
\end{eqnarray}

\begin{eqnarray}
\rho^0_{ss}\left(7\right) &=&\langle\frac{\alpha_{s}GG}{\pi}\rangle \Big\{\frac{m_c^3\langle\bar{q}q\rangle}{288\pi^2}\int_{x_i}^1 dx \frac{\left(1-x\right)^2}{x^3}\delta\left(s-\tilde{m}_c^2\right)+\frac{m_c\langle\bar{q}q\rangle} {288\pi^2}\int_{x_i}^1 dx \left(2-\frac{3\left(1-x\right)}{x^2}\right) \nonumber\\
&& +\frac{5m_s \langle\bar{s}s\rangle}{192\pi^2}\int_{x_i}^1 dx \left(1-x\right) \left[2+s \delta\left(s-\tilde{m}_c^2\right)\right] -\frac{m_s \langle\bar{s}s\rangle}{576\pi^2}\int_{x_i}^1 dx x \left[2+s\delta\left(s-\tilde{m}_c^2\right)\right]\Big\}\nonumber\, ,
\end{eqnarray}

\begin{eqnarray}
\rho^0_{ss}\left(8\right) &=&\frac{\langle\bar{s}s\rangle \langle\bar{s}g_s\sigma Gs\rangle}{48\pi^2} \int_{x_i}^1 dx \left(1-4x\right) \left[2+s\delta\left(s-\tilde{m}_c^2\right)\right] \nonumber\\
&&-\frac{m_s m_c \left(40\langle\bar{q}q\rangle\langle\bar{s}g_s\sigma Gs\rangle+39\langle\bar{s}s\rangle \langle\bar{q}g_s\sigma Gq\rangle \right)}{288\pi^2} \delta\left(s-m_c^2\right)\nonumber\\
&&+\frac{m_s m_c \langle\bar{q}q\rangle\langle\bar{s}g_s\sigma Gs\rangle}{24\pi^2} \int_{x_i}^1 dx \frac{1}{x} \delta\left(s-\tilde{m}_c^2\right)\nonumber\, ,
\end{eqnarray}

\begin{eqnarray}
\rho^0_{ss}\left(9\right) &=&-\frac{4m_c\langle\bar{q}q\rangle \langle\bar{s}s\rangle^2}{9} \delta\left(s-m_c^2\right)\nonumber\, ,
\end{eqnarray}

\begin{eqnarray}
\rho^0_{ss}\left(10\right) &=&\frac{\langle\bar{s}g_s\sigma Gs\rangle^2}{1536\pi^2} \left\{8\left(1+\frac{s}{T^2}\right)\delta\left(s-m_c^2\right)+3\int_{x_i}^1 dx \left(1+\frac{s}{T^2}\right) \delta\left(s-\tilde{m}_c^2\right)\right\}\nonumber\\
&&+\frac{m_s m_c \langle\bar{q}g_s\sigma Gq\rangle \langle\bar{s}g_s\sigma Gs\rangle}{192\pi^2} \left(-\frac{2}{T^2}+\frac{7s}{T^4}\right)\delta\left(s-m_c^2\right)\nonumber\\
&&+\langle\frac{\alpha_{s}GG}{\pi}\rangle \Big\{ \frac{\langle\bar{s}s\rangle^2}{216} \left[-3\int_{x_i}^1 dx \left(1+\frac{s}{T^2}\right) \delta\left(s-\tilde{m}_c^2\right)+2\left(1+\frac{s}{T^2}\right) \delta\left(s-m_c^2\right)\right] \nonumber\\
&&-\frac{m_s m_c^3 \langle\bar{q}q\rangle \langle\bar{s}s\rangle}{36T^4} \int_{x_i}^1 dx \frac{1}{x^3} \delta\left(s-\tilde{m}_c^2\right) -\frac{m_c^2\langle\bar{s}s\rangle^2}{108T^4}\int_{x_i}^1 dx \frac{s\left(1-x\right)}{x^2}  \delta\left(s-\tilde{m}_c^2\right)\nonumber\\
&&+\frac{m_s m_c \langle\bar{q}q\rangle \langle\bar{s}s\rangle}{12T^2} \int_{x_i}^1 dx \frac{1}{x^2} \delta\left(s-\tilde{m}_c^2\right)+\frac{m_s m_c \langle\bar{q}q\rangle\langle\bar{s}s\rangle}{432}\left(-\frac{3}{T^2}+\frac{14s}{T^4}\right) \delta\left(s-m_c^2\right) \Big\}\nonumber\, ,
\end{eqnarray}

\begin{eqnarray}
\rho^0_{ss}\left(11\right) &=&\frac{m_c^3 \langle\bar{s}s\rangle \left(2\langle\bar{q}q\rangle \langle\bar{s}g_s\sigma Gs\rangle+\langle\bar{s}s\rangle \langle\bar{q}g_s\sigma Gq\rangle\right)}{9T^4}\delta\left(s-m_c^2\right)\nonumber\\
&&-\frac{m_c \langle\bar{q}q\rangle \langle\bar{s}s\rangle \langle\bar{s}g_s\sigma Gs\rangle}{18T^2}\delta\left(s-m_c^2\right)\nonumber\, ,
\end{eqnarray}

\begin{eqnarray}
\rho^1_{ss}\left(0\right) &=& \frac{1}{1536\pi^6}\int_{x_i}^1 dx x\left(1-x\right)^4 s\left(s-\tilde{m}_c^2\right)^3 \nonumber\, ,
\end{eqnarray}

\begin{eqnarray}
\rho^1_{ss}\left(3\right) &=&-\frac{m_c \langle\bar{q}q\rangle}{288\pi^4} \int_{x_i}^1 dx \left[\left(4-x\right)\left(1-x\right)^2 \left(s-\tilde{m}_c^2\right)^2 +4\left(1-x\right)^3 s\left(s-\tilde{m}_c^2\right) \right]\nonumber\\
&&-\frac{m_s \langle\bar{s}s\rangle}{16\pi^4} \int_{x_i}^1 dx x\left(1-x\right)^2 \left(2s-\tilde{m}_c^2\right)\left(s-\tilde{m}_c^2\right) \nonumber\, ,
\end{eqnarray}

\begin{eqnarray}
\rho^1_{ss}\left(4\right) &=&-\langle\frac{\alpha_{s}GG}{\pi}\rangle \left\{ \frac{1}{4608\pi^4}\int_{x_i}^1 dx \left(1+2x\right)\left(1-x\right)^2 \left(s-\tilde{m}_c^2\right)^2 \nonumber +\frac{m_c^2}{4608\pi^4}\int_{x_i}^1 dx \frac{\left(1-x\right)^4}{x^2} s \right\}\nonumber\, ,
\end{eqnarray}

\begin{eqnarray}
\rho^1_{ss}\left(5\right) &=&-\frac{m_s\langle\bar{s}g_s\sigma Gs\rangle}{384\pi^4} \int_{x_i}^1 dx \left[2\left(1-19x\right) \left(1-x\right) s -\left(1-23x\right) \left(1-x\right) \tilde{m}_c^2 \right] \nonumber\\
&&-\frac{5m_c\langle\bar{q}g_s\sigma Gq\rangle}{1152\pi^4} \int_{x_i}^1 dx \left[\left(3x-5\right)\left(1-x\right)s -\left(x-3\right)\left(1-x\right) \tilde{m}_c^2 \right]\nonumber\, ,
\end{eqnarray}

\begin{eqnarray}
\rho^1_{ss}\left(6\right) &=&\frac{\langle\bar{s}s\rangle^2}{6\pi^2} \int_{x_i}^1 dx x\left(1-x\right) \left(2s-\tilde{m}_c^2\right)
\nonumber\\
&&-\frac{m_s m_c\langle\bar{q}q\rangle \langle\bar{s}s\rangle}{18\pi^2} \int_{x_i}^1 dx \left\{\left(1-x\right) \left[1+2s\delta\left(s-\tilde{m}_c^2\right)\right]-5\right\} \nonumber\, ,
\end{eqnarray}

\begin{eqnarray}
\rho^1_{ss}\left(7\right) &=&\langle\frac{\alpha_{s}GG}{\pi}\rangle \Bigg\{-\frac{m_c\langle\bar{q}q\rangle} {864\pi^2}\int_{x_i}^1 dx \left(-4+\frac{\left(3-2x\right)\left(1-x\right)}{x^2}\right)\nonumber\\
&&-\frac{3m_s \langle\bar{s}s\rangle}{576\pi^2}\int_{x_i}^1 dx \left[4x-3+s\left(2x-1\right) \delta\left(s-\tilde{m}_c^2\right)\right] \nonumber\\
&&+\frac{m_c^3\langle\bar{q}q\rangle}{2592\pi^2}\int_{x_i}^1 dx \left[\frac{\left(2+x\right)\left(1-x\right)^2}{x^3} +\frac{2s\left(1-x\right)^3}{x^3 T^2} \right] \delta\left(s-\tilde{m}_c^2\right) \nonumber\\
&&+\frac{m_s m_c^2 \langle\bar{s}s\rangle}{288\pi^2} \int_{x_i}^1 dx \frac{\left(1-x\right)^2}{x^2}\left(1+ \frac{s}{T^2}\right) \delta\left(s-\tilde{m}_c^2\right)\nonumber\\
&&-\frac{m_c\langle\bar{q}q\rangle}{1728\pi^2}\int_{x_i}^1 dx \frac{\left(5x^2-3x+2\right)\left(1-x\right)}{x^2} \left[1+2s\delta\left(s-\tilde{m}_c^2\right)\right] \Bigg\}\nonumber\, ,
\end{eqnarray}

\begin{eqnarray}
\rho^1_{ss}\left(8\right) &=&-\frac{\langle\bar{s}s\rangle \langle\bar{s}g_s\sigma Gs\rangle}{144\pi^2} \int_{x_i}^1 dx \left(12x-1\right) \left[1+s\delta\left(s-\tilde{m}_c^2\right)\right] \nonumber\\
&&-\frac{m_s m_c \left(64\langle\bar{q}q\rangle\langle\bar{s}g_s\sigma Gs\rangle+61\langle\bar{s}s\rangle \langle\bar{q}g_s\sigma Gq\rangle \right)}{864\pi^2} \delta\left(s-m_c^2\right)\nonumber\\
&&-\frac{m_s m_c  \left(8\langle\bar{q}q\rangle\langle\bar{s}g_s\sigma Gs\rangle+11\langle\bar{s}s\rangle \langle\bar{q}g_s\sigma Gq\rangle \right)}{864\pi^2} \int_{x_i}^1 dx \left(1-\frac{2s}{T^2}\right)\delta\left(s-\tilde{m}_c^2\right) \nonumber\\
&&+\frac{m_s m_c \langle\bar{q}q\rangle\langle\bar{s}g_s\sigma Gs\rangle}{144\pi^2} \int_{x_i}^1 dx \frac{1}{x} \delta\left(s-\tilde{m}_c^2\right)\nonumber\, ,
\end{eqnarray}

\begin{eqnarray}
\rho^1_{ss}\left(9\right) &=&-\frac{2m_c\langle\bar{q}q\rangle \langle\bar{s}s\rangle^2}{9} \delta\left(s-m_c^2\right)\nonumber\, ,
\end{eqnarray}

\begin{eqnarray}
\rho^1_{ss}\left(10\right) &=&-\frac{\langle\bar{s}g_s\sigma Gs\rangle^2}{4608\pi^2} \left[-\frac{40s}{T^2}\delta\left(s-m_c^2\right)+19\int_{x_i}^1 dx \frac{s}{T^2} \delta\left(s-\tilde{m}_c^2\right) \right] \nonumber\\
&&+\frac{m_s m_c \langle\bar{q}g_s\sigma Gq\rangle \langle\bar{s}g_s\sigma Gs\rangle}{1728\pi^2}\left(\frac{8}{T^2}+\frac{25s}{T^4}\right) \delta\left(s-m_c^2\right)\nonumber\\
&&+\langle\frac{\alpha_{s}GG}{\pi}\rangle \left\{ -\frac{\langle\bar{s}s\rangle^2}{216} \left[\int_{x_i}^1 dx\left(2+\frac{s}{T^2}\right) \delta\left(s-\tilde{m}_c^2\right)-\frac{2s}{T^2} \delta\left(s-m_c^2\right)\right] \right.\nonumber\\
&&+\frac{m_s m_c \langle\bar{q}q\rangle\langle\bar{s}s\rangle}{1296}\left(\frac{5}{T^2}+\frac{18s}{T^4}\right)\delta\left(s-m_c^2\right) \nonumber\\
&&-\frac{m_s m_c^3 \langle\bar{q}q\rangle \langle\bar{s}s\rangle}{324} \int_{x_i}^1 dx \left(\frac{5\left(2-x\right)}{x^3 T^4}-\frac{2s\left(1-x\right) }{x^3 T^6} \right) \delta\left(s-\tilde{m}_c^2\right)\nonumber\\
&&-\frac{m_s m_c \langle\bar{q}q\rangle \langle\bar{s}s\rangle}{1296} \int_{x_i}^1 dx \left(-\frac{3\left(32-11x\right)}{x^2 T^2}+\frac{2s\left(12-11x\right)}{x^2 T^4} \right) \delta\left(s-\tilde{m}_c^2\right)\nonumber\\
&&\left.-\frac{m_c^2\langle\bar{s}s\rangle^2}{108}\int_{x_i}^1 dx \frac{1-x}{x^2} \left(-\frac{1}{T^2}+\frac{s}{T^4}\right) \delta\left(s-\tilde{m}_c^2\right) \right\}\nonumber\, ,
\end{eqnarray}

\begin{eqnarray}
\rho^1_{ss}\left(11\right) &=&\frac{m_c^3 \langle\bar{s}s\rangle \left(2\langle\bar{q}q\rangle \langle\bar{s}g_s\sigma Gs\rangle+\langle\bar{s}s\rangle \langle\bar{q}g_s\sigma Gq\rangle\right)}{18T^4}\delta\left(s-m_c^2\right)\nonumber\\
&&-\frac{m_c \langle\bar{q}q\rangle \langle\bar{s}s\rangle \langle\bar{s}g_s\sigma Gs\rangle}{108T^2}\delta\left(s-m_c^2\right)\nonumber\, ,
\end{eqnarray}

\begin{eqnarray}
\rho^2_{ss}\left(0\right) &=&\frac{1}{38400\pi^6}\int_{x_i}^1 dx x\left(1-x\right)^5 \left[\left(s-\tilde{m}_c^2\right)^4 +16s\left(s-\tilde{m}_c^2\right)^3+16s^2\left(s-\tilde{m}_c^2\right)^2\right] \nonumber\\
&&+\frac{1}{11520\pi^6}\int_{x_i}^1 dx x\left(1-x\right)^4 \left(9s-2\tilde{m}_c^2\right) \left(s-\tilde{m}_c^2\right)^3\nonumber\, ,
\end{eqnarray}

\begin{eqnarray}
\rho^2_{ss}\left(3\right) &=&-\frac{m_s \langle\bar{s}s\rangle}{360\pi^4} \int_{x_i}^1 dx \left\{3x\left(1-x\right)^2 \left(34s-11\tilde{m}_c^2\right) \left(s-\tilde{m}_c^2\right)\right.\nonumber\\
&&\left.-x\left(1-x\right)^3 \left[3\left(s-\tilde{m}_c^2\right)^2 +24s\left(s-\tilde{m}_c^2\right)+8s^2\right] \right\}\nonumber\\
&&-\frac{m_c\langle\bar{q}q\rangle}{144\pi^4} \int_{x_i}^1 dx \left[3\left(1-x\right)^2 \left(s-\tilde{m}_c^2\right)^2 +\left(1-x\right)^3 \left(5s-\tilde{m}_c^2\right) \left(s-\tilde{m}_c^2\right) \right]\nonumber\, ,
\end{eqnarray}

\begin{eqnarray}
\rho^2_{ss}\left(4\right) &=&\langle\frac{\alpha_{s}GG}{\pi}\rangle \Big\{ -\frac{1}{34560\pi^4}\int_{x_i}^1 dx \left(1+2x\right) \left(1-x\right)^2 \left(113s-37\tilde{m}_c^2\right)\left(s-\tilde{m}_c^2\right)\nonumber \\
&&+\frac{1}{69120\pi^4}\int_{x_i}^1 dx \left(1+3x\right) \left(1-x\right)^3 \left[3\left(s-\tilde{m}_c^2\right)^2 +24s\left(s-\tilde{m}_c^2\right)+8s^2\right]\nonumber\\
&&-\frac{m_c^2}{43200\pi^4}\int_{x_i}^1 dx \frac{\left(1-x\right)^5}{x^2} \left[3\left(7s-5\tilde{m}_c^2\right)+2s^2 \delta\left(s-\tilde{m}_c^2\right)\right] \nonumber\\
&&-\frac{m_c^2}{34560\pi^4}\int_{x_i}^1 dx \frac{\left(1-x\right)^4}{x^2} \left(15s-8\tilde{m}_c^2\right) \Big\}\nonumber\, ,
\end{eqnarray}

\begin{eqnarray}
\rho^2_{ss}\left(5\right) &=&\frac{7m_c\langle\bar{q}g_s\sigma Gq\rangle}{288\pi^4} \int_{x_i}^1 dx \left[\left(1-x\right) \left(s-\tilde{m}_c^2\right) +\left(1-x\right)^2 \left(3s-\tilde{m}_c^2\right) \right]\nonumber\\
&&+\frac{m_s\langle\bar{s}g_s\sigma Gs\rangle}{5760\pi^4} \int_{x_i}^1 dx \big\{x\left(1-x\right) \left(1320s-652\tilde{m}_c^2\right)
\nonumber\\
&&-16x\left(1-x\right)^2 \left[3\left(5s-\tilde{m}_c^2\right)+4s^2\delta\left(s-\tilde{m}_c^2\right)\right] +30\left(1-x\right)^2\left(2s-\tilde{m}_c^2\right) \big\}\nonumber\, ,
\end{eqnarray}

\begin{eqnarray}
\rho^2_{ss}\left(6\right) &=&-\frac{m_s m_c\langle\bar{q}q\rangle \langle\bar{s}s\rangle}{9\pi^2} \int_{x_i}^1 dx \left\{\left(1-x\right) \left[1+2s\delta\left(s-\tilde{m}_c^2\right)\right]-5\right\} \nonumber\\
&& +\frac{\langle\bar{s}s\rangle^2}{3\pi^2} \int_{x_i}^1 dx x\left(1-x\right) \left(2s-\tilde{m}_c^2\right)\nonumber\, ,
\end{eqnarray}

\begin{eqnarray}
\rho^2_{ss}\left(7\right) &=&\langle\frac{\alpha_{s}GG}{\pi}\rangle \left\{\frac{m_c^3\langle\bar{q}q\rangle}{1296\pi^2}\int_{x_i}^1 dx \left[\frac{\left(2+x\right)\left(1-x\right)^2}{x^3} +\frac{2s\left(1-x\right)^3}{x^3 T^2} \right] \delta\left(s-\tilde{m}_c^2\right) \right.\nonumber\\
&&-\frac{m_c\langle\bar{q}q\rangle} {864\pi^2}\int_{x_i}^1 dx \left[-8+\frac{\left(7x^2-13x+8\right) \left(1-x\right)}{x^2}+\frac{2s\left(7x^2-5x+2\right) \left(1-x\right)}{x^2} \delta\left(s-\tilde{m}_c^2\right) \right] \nonumber\\
&&-\frac{m_s \langle\bar{s}s\rangle}{8640\pi^2}\int_{x_i}^1 dx \Big\{150x\left[1+s\delta\left(s-\tilde{m}_c^2\right)\right] +2\left(1-x\right)\left[-121+s\delta\left(s-\tilde{m}_c^2\right)\right]\nonumber\\
&&-4\left(1-x\right)^2 \left[3+4\left(s+\frac{s^2}{T^2}\right)\delta\left(s-\tilde{m}_c^2\right) \right]  \Big\}\nonumber\\
&&\left. -\frac{m_s m_c^2 \langle\bar{s}s\rangle}{6480\pi^2} \int_{x_i}^1 dx \frac{\left(1-x\right)^2}{x^2}\left[\left(1+2x\right) + \frac{s\left(8x-77\right)}{T^2} +\frac{8s^2\left(1-x\right)}{T^4}  \right] \delta\left(s-\tilde{m}_c^2\right) \right\}\nonumber\, ,
\end{eqnarray}

\begin{eqnarray}
\rho^2_{ss}\left(8\right) &=&-\frac{\langle\bar{s}s\rangle \langle\bar{s}g_s\sigma Gs\rangle}{72\pi^2} \int_{x_i}^1 dx \left(12x+1\right) \left[1+s\delta\left(s-\tilde{m}_c^2\right)\right] \nonumber\\
&&-\frac{m_s m_c \left(64\langle\bar{q}q\rangle\langle\bar{s}g_s\sigma Gs\rangle+59\langle\bar{s}s\rangle \langle\bar{q}g_s\sigma Gq\rangle \right)}{432\pi^2} \delta\left(s-m_c^2\right)\nonumber\\
&&-\frac{m_s m_c  \left(8\langle\bar{q}q\rangle\langle\bar{s}g_s\sigma Gs\rangle+13\langle\bar{s}s\rangle \langle\bar{q}g_s\sigma Gq\rangle \right)}{432\pi^2} \int_{x_i}^1 dx \left(1-\frac{2s}{T^2}\right)\delta\left(s-\tilde{m}_c^2\right) \nonumber\\
&&-\frac{m_s m_c \langle\bar{q}q\rangle\langle\bar{s}g_s\sigma Gs\rangle}{72\pi^2} \int_{x_i}^1 dx \frac{1}{x} \delta\left(s-\tilde{m}_c^2\right)\nonumber\, ,
\end{eqnarray}

\begin{eqnarray}
\rho^2_{ss}\left(9\right) &=&-\frac{4m_c\langle\bar{q}q\rangle \langle\bar{s}s\rangle^2}{9} \delta\left(s-m_c^2\right)\nonumber\, ,
\end{eqnarray}

\begin{eqnarray}
\rho^2_{ss}\left(10\right) &=&\frac{7\langle\bar{s}g_s\sigma Gs\rangle^2}{2304\pi^2} \left[\frac{8s}{T^2}\delta\left(s-m_c^2\right)+\frac{5}{3}\int_{x_i}^1 dx \frac{s}{T^2} \delta\left(s-\tilde{m}_c^2\right) \right] \nonumber\\
&&+\frac{m_s m_c \langle\bar{q}g_s\sigma Gq\rangle \langle\bar{s}g_s\sigma Gs\rangle}{864\pi^2}\left(\frac{16}{T^2}+\frac{23s}{T^4}\right) \delta\left(s-m_c^2\right)\nonumber\\
&&+\langle\frac{\alpha_{s}GG}{\pi}\rangle \left\{ -\frac{\langle\bar{s}s\rangle^2}{108} \left[\int_{x_i}^1 dx\left(2+\frac{s}{T^2}\right) \delta\left(s-\tilde{m}_c^2\right)-\frac{2s}{T^2} \delta\left(s-m_c^2\right)\right] \right.\nonumber\\
&&+\frac{m_s m_c \langle\bar{q}q\rangle\langle\bar{s}s\rangle}{648}\left(\frac{7}{T^2}+\frac{18s}{T^4}\right)\delta\left(s-m_c^2\right) \nonumber\\
&&-\frac{m_s m_c^3 \langle\bar{q}q\rangle \langle\bar{s}s\rangle}{162} \int_{x_i}^1 dx \left(\frac{5\left(2-x\right)}{x^3 T^4}-\frac{2s\left(1-x\right) }{x^3 T^6} \right) \delta\left(s-\tilde{m}_c^2\right)\nonumber\\
&&-\frac{m_s m_c \langle\bar{q}q\rangle \langle\bar{s}s\rangle}{648} \int_{x_i}^1 dx \left(\frac{3\left(13x-32\right)}{x^2 T^2}+\frac{2s\left(12-13x\right)}{x^2 T^4} \right) \delta\left(s-\tilde{m}_c^2\right)\nonumber\\
&&\left.-\frac{m_c^2\langle\bar{s}s\rangle^2}{54}\int_{x_i}^1 dx \frac{1-x}{x^2} \left(-\frac{1}{T^2}+\frac{s}{T^4}\right) \delta\left(s-\tilde{m}_c^2\right) \right\}\nonumber\, ,
\end{eqnarray}

\begin{eqnarray}
\rho^2_{ss}\left(11\right) &=&\frac{m_c^3 \langle\bar{s}s\rangle \left(2\langle\bar{q}q\rangle \langle\bar{s}g_s\sigma Gs\rangle+\langle\bar{s}s\rangle \langle\bar{q}g_s\sigma Gq\rangle\right)}{9T^4}\delta\left(s-m_c^2\right)\nonumber\\
&&+\frac{m_c \langle\bar{q}q\rangle \langle\bar{s}s\rangle \langle\bar{s}g_s\sigma Gs\rangle}{54T^2}\delta\left(s-m_c^2\right)\nonumber\, ,
\end{eqnarray}
where $\tilde{m}_c^2=\frac{m_c^2}{x}$ and $x_i=\frac{m_c^2}{s}$.

\section*{Acknowledgements}
This work is supported by National Natural Science Foundation, Grant Number  12175068.

\end{document}